\documentclass[12pt]{article}
\usepackage{amsmath,amssymb,amsfonts}
\usepackage[dvips]{graphicx}
\usepackage{epsfig}

\usepackage[vcentermath]{youngtab}

\makeatletter \@addtoreset{equation}{section} \makeatother

\renewcommand{\baselinestretch}{1.2}
\def\det{{\rm det}}

\newcommand{\be}{\begin{eqnarray}}
\newcommand{\ee}{\end{eqnarray}}
\newcommand{\nn}{\nonumber}
\newcommand{\bn}{\begin{enumerate}}
\newcommand{\en}{\end{enumerate}}

\begin{document}

\renewcommand{\thefootnote}{\alph{footnote}}

\begin{titlepage}

\begin{center}
\hfill {\tt KIAS-P12038}\\
\hfill {\tt SNUTP12-002}\\

\vspace{2cm}

{\Large\bf M5-branes from gauge theories on the 5-sphere}

\vspace{2cm}

\renewcommand{\thefootnote}{\alph{footnote}}

{\large Hee-Cheol Kim$^1$ and Seok Kim$^2$}

\vspace{1cm}

\textit{ $^1$School of Physics, Korea Institute for Advanced Study,
Seoul 130-012, Korea.}\\

\vspace{0.2cm}

\textit{$^2$Department of Physics and Astronomy \& Center for
Theoretical Physics,\\
Seoul National University, Seoul 151-747, Korea.}\\

\vspace{1cm}

E-mails: {\tt heecheol1@gmail.com, 
skim@phya.snu.ac.kr}

\end{center}

\vspace{2cm}

\begin{abstract}

We use the 5-sphere partition functions of supersymmetric Yang-Mills theories to
explore the $(2,0)$ superconformal theory on $S^5\!\times\!S^1$. The 5d theories can be
regarded as Scherk-Schwarz reductions of the 6d theory along the circle. In a special limit,
the perturbative partition function takes the form of the Chern-Simons partition function
on $S^3$. With a simple non-perturbative completion, it becomes a 6d index which
captures the degeneracy of a sector of BPS states as well as the index version of the
vacuum Casimir energy. The Casimir energy exhibits the $N^3$ scaling at large $N$.
The large $N$ index for $U(N)$ gauge group also completely agrees with the supergravity
index on $AdS_7\times S^4$.

\end{abstract}

\end{titlepage}

\renewcommand{\thefootnote}{\arabic{footnote}}

\setcounter{footnote}{0}

\renewcommand{\baselinestretch}{1}

\tableofcontents

\renewcommand{\baselinestretch}{1.2}

\section{Introduction}

M5-brane is one of the most mysterious objects in M-theory \cite{Hull:1994ys}.
M2- and M5-branes, which are two important ingredients of M-theory, are known to support
strange numbers of light degrees of freedom on their worldvolumes \cite{Klebanov:1996un}.
Although the $N^{3/2}$ scalings for $N$ coincident M2-branes have been recently understood
in some detail \cite{Aharony:2008ug,Drukker:2010nc}, the $N^3$
scalings for $N$ M5-branes are not very solidly understood in a microscopic way.

M-theory is related to 10d string theories by having an extra direction emerging in strongly
coupled string theories \cite{Hull:1994ys}, being a circle for the type IIA strings.
This relation is mainly supported by identifying D0-brane states with the Kaluza-Klein states
of M-theory along the circle. Such a relation could still hold in
Euclidean type IIA/M-theories on various curved manifolds with a circle factor.

The relation between type IIA/M-theories via a circle compactification also yields
a similar relation between the D4-brane and M5-brane theories. On M5-branes probing
flat transverse space or its $\mathbb{Z}_2$ orbifold, there live 6d $(2,0)$ superconformal
theories associated with $A_n$ or $D_n$ type gauge groups. The full set of known
6d $(2,0)$ theories actually come in an $ADE$ classification \cite{Witten:1995zh}.
The microscopic details of these
theories are largely unknown. Dimensional reductions of these 6d theories along a small
circle admit descriptions by 5d maximally supersymmetric Yang-Mills theories.
Naively, the resulting 5d theory is supposed to be a dimensional reduction, after which
one expects that information on the 6d physics is lost. There appeared some evidence that
careful studies of the strong-coupling or non-perturbative physics of the 5d theory let us
extract the nontrivial information on the 6d theory compactified on the circle \cite{Aharony:1997th,Douglas:2010iu}. In Minkowskian dynamics, crucial roles are played by
the instanton solitons in the 5d theory, similar to the way in which type IIA D0-branes are
crucial for reconstructing the KK states of the extra circle. In particular, in BPS sectors,
it has been shown in detail that the instanton partition function yields various (expected or
novel) results for 6d $(2,0)$ theory compactified on a circle \cite{Kim:2011mv}: this includes the rigorous proof of the uniqueness of $U(1)$ multi-instanton bound states, discovery of
novel self-dual string bound states which explains some enhancements of degrees of freedom in the Coulomb branch, the study of the symmetric phase instanton index and its agreement with
the DLCQ gravity dual index on $AdS_7\times S^4$.

In this paper, we apply the same idea to the 6d theory on $S^5\times S^1$, and study
them from 5d gauge theories on $S^5$. As the 5d gauge theories (at least apparently) look non-renormalizable, there is a general issue on how to make quantum calculations sensible.
There appeared proposals on possible finiteness of maximally supersymmetric theories in
5d \cite{Douglas:2010iu}. (See also \cite{Lambert:2010wm} for an earlier work.) But even
if this is true, having a good control over all the 5d quantum fluctuations would be
generally difficult. Just as those considered in \cite{Kim:2011mv}, there are many supersymmetric observables which rely less sensitively on quantum fluctuations.
We expect that the BPS observables that we consider in this paper would also be safe:
in fact, based on localization, we are led to consider a supersymmetric path integral
which is secretly Gaussian, for which the UV divergence issue is almost trivial.
So we base our studies on a much more modest but solidly testable proposal that
5d supersymmetric Yang-Mills theory describes 6d $(2,0)$ theory compactified on a
circle at least in the BPS sector. Note that this proposal is not necessarily restricted
to maximal SYM: although we focus on maximal SYM in this paper, we generalize the
study to less supersymmetric theories in a follow up work \cite{Kim:2012qf}.

The $(2,0)$ theory on $S^5\times S^1$ is interesting for various reasons.
Firstly, any 6d CFT on flat spacetime can be put on $S^5\times\mathbb{R}$ by radial
quantization, where $\mathbb{R}$ is the (Euclidean) time direction. Depending on how one
compactifies the time direction to a circle, the resulting partition function will be an
appropriate index which counts BPS states of this theory. In particular, $S^5\times\mathbb{R}$ is the conformal boundary of global $AdS_7$, so that the large $N$ limits (if available)
of these theories could have gravity duals on global $AdS_7$ \cite{Maldacena:1997re}.
AdS$_7$/CFT$_6$ is perhaps the least understood duality among various AdS/CFT proposals,
on which we can shed lights with our studies.

When the circle size is small, we are naturally led to study the Euclidean
supersymmetric Yang-Mills theory on the 5-sphere. For the $ADE$ cases, we study the 5d
gauge theories with corresponding gauge groups. For $A_n$ and $D_n$ cases, they can be
understood intuitively as living on `Euclidean D4-branes' wrapping the 5-sphere, if
one reduces the 6d theory on the circle interpreted as the M-theory circle.

We first construct and calculate the partition function of a Yang-Mills quantum field theory
on $S^5$ preserving $16$ real SUSY. To motivate the construction from the 6d $(2,0)$ theory,
we first consider the Abelian 6d $(2,0)$ theory. As this free theory on $\mathbb{R}^6$ is
conformal, one can radially quantize it to obtain a theory on $S^5\times\mathbb{R}$.
The $32$ Killing spinors satisfy one of the two Killing spinor equations:
\begin{equation}
  \nabla_M\epsilon=\pm\frac{1}{2r}\Gamma_M\Gamma_\tau\epsilon\ ,
\end{equation}
where $r$ is the radius of $S^5$ and $\tau$ is the Euclidean time.
Since the dependence of $\epsilon$ on $\tau$ is $e^{\pm\frac{1}{2r}\tau}$,
one cannot naively compactify this theory preserving all $32$ SUSY. Instead,
one can introduce an R-symmetry twist (or a Scherk-Schwarz reduction) to obtain a theory
on $S^5\times S^1$ with as much as $16$ SUSY.
This can be done by picking an $SO(2)\subset SO(5)$ R-symmetry. The resulting theory after
the 5d reduction, with tensor-vector dualization, can be straightforwardly generalized to
non-Abelian theories with arbitrary gauge group. Due to the R-symmetry twist, the maximal SYM
on $S^5$ preserves only $SO(2)\times SO(3)$ part of $SO(5)$ R-symmetry.

We calculate and study the partition function of this maximal SYM on $S^5$. We employ
the localization technique to obtain the perturbative contribution given by a simple matrix
integral. We also suggest a simple non-perturbative correction, which is proved in
a follow-up paper \cite{Kim:2012qf}. The M-theory interpretation demands
us to relate the 5d gauge coupling $g_{YM}$ and the circle radius $r_1$ as
\begin{equation}
  \frac{4\pi^2}{g_{YM}^2}=\frac{1}{r_1}=\frac{2\pi}{r\beta}\ ,
\end{equation}
where $\beta$ is the (dimensionless) inverse `temperature' like chemical potential. In flat
Minkowskian space, this is relating the instanton (or D0-brane) mass with the Kaluza-Klein
mass on the extra circle. With this interpretation, and also with the R-symmetry twist on
which we elaborate in
section 2, the 5-sphere partition function is identified as an index of the 6d theory with
the chemical potential $\beta$. This index counts BPS states on $S^5\times\mathbb{R}$,
or local BPS operators on $\mathbb{R}^6$.
The fact that our 5d partition function takes the form of an index, with all coefficients
being integers when expanded in the fugacity $e^{-\beta}$, strongly supports that
the 5d Yang-Mills theory is nontrivially capturing the 6d physics.

In the later part of this paper, we mostly consider the $U(N)$ gauge theory in 5d, to
study the $A_{N-1}$ type $(2,0)$ theory in 6d times a decoupled free sector. However, we
comment on some important general features for all $ADE$ gauge groups, and also on possible
fate of the theories with non-$ADE$ gauge groups, including $BCFG$.

Our partition function captures two different features of the 6d theory. Firstly, it tells us
the degeneracy information of the BPS states of the 6d theory.  Secondly, and perhaps more
interestingly, it contains the information on the 6d vacuum on $S^5\times\mathbb{R}$.
The unique
vacuum of the radially quantized 6d theory has nonzero Casimir energy. In the large $N$ limit
of the $SU(N)$ and $SO(2N)$ cases, the $AdS_7$ gravity dual predicts its value to be nonzero
and proportional to $N^3$ \cite{Awad:2000aj}. From the gravity side,
this is basically the same $N^3$ appearing in all $AdS_7$ gravity calculations, coming from
$\frac{\ell^5}{G_7}$ combination of the $AdS_7$ radius $\ell$ and 7d Newton constant $G_7$.
Our partition function captures the `index version' of the vacuum Casimir energy, which also
exhibits the $N^3$ scaling in the large $N$ limit. See section 3 and appendix B for what we
mean by the `index Casimir energy.' The difference between the normal Casimir energy of
CFT and ours is that ours uses an unconventional regularization for the Casimir energy,
which is naturally chosen by the definition of the index we consider.

Curiously, the perturbative partition functions of our theories with $16$ SUSY on $S^5$
turn out to take identical forms as the partition functions of pure Chern-Simons theories
on $S^3$, when we appropriately identify the Chern-Simons coupling constant with the 5d
coupling constant.

Upon adding a simple non-perturbative correction to the above perturbative part,
we also show that the $U(N)$ index completely agrees with the supergravity index
on $AdS_7\times S^4$ in the large $N$ limit. Also, our finite $N$ index is a function
which appears in various different physical/mathematical contexts. See section 3.2 for
the details.

We also provide a matrix integral form of the perturbative part of a generalized
partition function, which we suppose to be a more refined 6d index with two chemical
potentials. For this we study a SYM theory on $S^5$ preserving $8$ SUSY, which can be
regarded as a Scherk-Schwarz reduction of the 6d $(2,0)$ theory with more general
$U(1)\subset SO(5)$ embedding. In one limit, we suggest that the generalized partition
function captures the spectrum of half-BPS states of the 6d theory, whose
general structures are explored, for instance, in \cite{Bhattacharyya:2007sa}.

The remaining part of this paper is organized as follows. In section 2, we motivate
our theory on $S^5$ by taking a Scherk-Schwarz reduction of the Abelian
6d $(2,0)$ theory. The resulting 5d theory is generalized to a non-Abelian theory on $S^5$.
In section 3, we calculate the perturbative partition function and show that
it takes the same form as the Chern-Simons partition function on $S^3$. Adding non-perturbative
corrections, we study the index Casimir energy, the large $N$ index and the dual gravity
index. We finally present a matrix integral form of a
generalized partition function which we expect to be a more refined 6d index.
Appendix A explains the scalar/spinor/vector spherical harmonics on $S^5$,
as well as some path integral calculations. Appendix B explains that the superconformal
indices (of which our partition function is a special sort) in various dimensions capture
the index version of Casimir energies and study their properties.

As we were finalizing the preparation of this manuscript, we received
\cite{Kallen-Qiu-Zabzine} which partly overlaps with our section 3.3. Their result is a
special case of ours in section 3.3 with $\Delta=\frac{1}{2}$.

\section{Maximal SYM on the 5-sphere}

\subsection{Motivation from Abelian theories}

As a motivation, we would like to reduce the radially quantized Abelian
$(2,0)$ theory on a circle to obtain a theory on $S^5$ with $16$ SUSY.
The resulting 5d theory will be generalized to non-Abelian
theories in section 2.2.

The $32$ Killing spinors on Minkowskian $S^5\times\mathbb{R}$ satisfy one of
the two equations
\begin{equation}\label{killing}
  \nabla_M\epsilon_\pm=\pm\frac{i}{2r}\Gamma_M\Gamma_0\epsilon_\pm\ ,
\end{equation}
where $M=0,1,2,3,4,5$, and $r$ is the radius of $S^5$. Taking $M=0$, one
finds the time dependence
\begin{equation}
  \epsilon_\pm(\tau)=e^{\mp\frac{i}{2r}t}\epsilon_{0\pm}\ .
\end{equation}
The spinors with two signs yield Poincare/conformal supercharges, respectively,
which should be suitably complex conjugate to each other.

We first consider the properties of our spinors in some detail. The matter and
Killing spinors of the 6d $(2,0)$ theory are all spinors in spacetime $SO(5,1)$
(or $SO(6)$ in Euclidean theories) and the $SO(5)_R$ R-symmetry.
The $8\times 8$ gamma matrices in 6d can be written in terms of the $4\times 4$
5d gamma matrices $\gamma_\mu$ (which shall be useful after a circle reduction) as
\begin{equation}
  \Gamma_\mu=\gamma_\mu\otimes\sigma_1\ ,\ \ \Gamma_\tau={\bf 1}_4\otimes\sigma_2
\end{equation}
on a Euclidean space. Multiplication of factor $i$ to $\Gamma_\tau$ will convert
it to the Lorentzian gamma matrices.
The 6d chirality matrix $\Gamma^{123456}=i\sigma_3$ demands that a chiral spinor
have $\sigma_3=+1$ eigenvalue. To be concrete, we take
the following representation of the 5d gamma matrices in this paper ($\sigma^{1,2,3}$
are Pauli matrices):
\begin{equation}
  \gamma^{1,2,3}=\sigma^{1,2,3}\otimes\sigma^{1}\ ,\ \
  \gamma^4={\rm 1}_2\otimes\sigma^2\ ,\ \ \gamma^5=-{\bf 1}_2\otimes\sigma^3\ .
\end{equation}
These satisfy $\gamma^{12345}=1$. Also, for the internal $SO(5)$ spinors, we
introduce the $4\times 4$ gamma matrices $\hat\gamma^I$ ($I=1,2,3,4,5$) as
\begin{equation}\label{internal-gamma}
  \hat\gamma^1=\sigma^1\otimes\sigma^1\ ,\ \ \hat\gamma^2=\sigma^2\otimes\sigma^1\ ,\ \
  \hat\gamma^4=\sigma^3\otimes\sigma^1\ ,\ \ \hat\gamma^5={\bf 1}_2\otimes\sigma^2\ ,\ \
  \hat\gamma^3=\hat\gamma^{1245}=-{\bf 1}_2\otimes\sigma^3\ ,
\end{equation}
which satisfy $\hat\gamma^{12345}=1$.

With the above convention for gamma matrices, one finds (in the Lorentzian case)
\begin{equation}
  (\Gamma_M)^T=\left(\Gamma_1,-\Gamma_2,\Gamma_3,-\Gamma_4,\Gamma_5,-\Gamma_0\right)
  =\pm C_\pm\Gamma_MC_\pm^{-1}
\end{equation}
with $C_+\sim\Gamma_{135}\sim\gamma_{24}\otimes\sigma^1\equiv C\otimes\sigma^1$ and
$C_-\sim\Gamma_{240}\sim\gamma_{24}\otimes\sigma_2=C\otimes\sigma^2$.
Here, $C$ is the charge conjugation matrix in 5d in our convention.
Killing spinors $\epsilon_\pm$ are
related by a symplectic charge conjugation, using either of $C_\pm$ together with
the $SO(5)_R\sim Sp(4)$ internal charge conjugation $\hat{C}\sim
\hat\gamma^{25}=i\sigma^2\otimes\sigma^3$. Namely, the Killing spinors satisfy
$\epsilon_-^T=\bar\epsilon_+C\otimes\hat{C}$. With the appearance of $\Gamma^0$ in
$\bar\epsilon_+=\epsilon_+^\dag\Gamma^0$, the symplectic charge conjugation with
Lorentzian signature does not flip the 6d chirality. Also, it is easy to see that the
equations (\ref{killing}) for $\epsilon_\pm$ correctly transform into each other by
the above conjugation. So $\epsilon_\pm$ can both be taken to be in the ${\bf 4}$
representation of $SO(6)$, yielding 6d $(2,0)$ SUSY.

On the other hand, in Euclidean 6d, one finds
\begin{equation}
  (\Gamma_M)^\ast=\left(\Gamma_1,-\Gamma_2,\Gamma_3,-\Gamma_4,\Gamma_5,-\Gamma_6
  \right)=\pm C_\pm\Gamma_MC_\pm^{-1}
\end{equation}
with same $C_\pm$ as in the Lorentzian case. So one may be tempted to relate $\epsilon_\pm$
by a similar symplectic Majorana condition $\epsilon_-\stackrel{?}{=}C\otimes\hat{C}\epsilon_+^\ast$.
This time, the charge conjugation flips the 6d chirality. Also, changing
$\Gamma_0$ on the right hand side of (\ref{killing}) to make it into $\Gamma_6$ along
$\tau$ direction, $\epsilon_\pm$ equations are no longer related to each other with the above
conjugation. A natural charge conjugation in the radially quantized Euclidean CFT is to accompany
it with the sign flip of $\tau$ \cite{Bhattacharyya:2007sa}, as this is changing particles
into anti-particles. (This is basically remembering the Lorentzian physics via $\tau=it$.)
Also, we multiply $\Gamma_6$ on the charge conjugation matrix to have all matter and Killing
spinors to have same chirality. Combining the charge conjugation with $\tau\rightarrow-\tau$
and a multiplication of $\Gamma_6$, one finds that the Euclidean version of (\ref{killing})
for $\epsilon_\pm$ are related to each other.
Thus, we have $32$ real Killing spinors in both Lorentzian and Euclidean
6d theories, all being chiral.

Now we consider the Euclidean theory with time $\tau$.
Since all Killing spinors depend on $\tau$, naive compactification
on $S^5\times S^1$ breaks all SUSY. To preserve $16$ SUSY, one suitably twists
the theory with an $SO(2)\subset SO(5)$ chemical potential to admit constant spinors
on $S^1$. Namely, taking a ${\bf 5}\rightarrow({\bf 3},{\bf 1})+({\bf 1},{\bf 2})$ decomposition of an $SO(5)\supset SO(3)\times SO(2)$ vector, one takes the $SO(2)$
which rotates ${\bf 2}$ and introduces
the background gauge field which covariantizes
\begin{equation}
  \nabla_\tau\rightarrow\ \nabla_\tau+\frac{i}{2r}\hat{\gamma}^{45}\ .
\end{equation}
This will correspond to introducing a chemical potential for the $SO(2)$ R-charge
of the 6d theory, which we shall explain in detail shortly. The $M=6$ components of
the Killing spinor equation then becomes
\begin{equation}
  \partial_\tau\epsilon_\pm=\frac{1}{2r}\left(\pm 1-i\hat\gamma^{45}\right)\epsilon_\pm\ .
\end{equation}
So in the case with $\pm$ sign, we take the Killing spinors with $i\hat{\gamma}^{45}=\pm 1$
eigenvalue to obtain $16$ SUSY. The resulting 5d Killing spinors satisfy
\begin{equation}
  \nabla_\mu\epsilon_\pm=\mp\frac{1}{2r}\Gamma_\mu\Gamma_\tau\epsilon_\pm
  =-\frac{i}{2r}\Gamma_\mu\Gamma_\tau\hat{\gamma}^{45}\epsilon_\pm\ .
\end{equation}
The 5d Killing spinor equation is
thus given by (using $\sigma^3\epsilon_\pm=\epsilon_\pm$)
\begin{equation}\label{killing-5d}
  \nabla_\mu\epsilon=\frac{1}{2r}\gamma_\mu\hat{\gamma}^{45}\epsilon\ ,
\end{equation}
which includes both $\epsilon_\pm$ cases.
This is the same as one of the Killing spinor equations studied in \cite{Blau:2000xg}
in 5d (although \cite{Blau:2000xg} discussed Minkowskian Einstein manifold).
In the reduced 5d perspective, we simply take the charge conjugation
$\epsilon_-=C\otimes\hat{C}\epsilon_+^\ast$ without knowing about $\tau$ flip. We
also forget the $\Gamma_6={\bf 1}\otimes\sigma_2$ multiplication by regarding
$\epsilon_\pm$ as 4 component spinors in 5d. $i\hat\gamma^{45}$ transforms under
this 5d charge conjugation as
\begin{equation}
  \hat{C}^{-1}(i\hat\gamma^{45})\hat{C}=-(i\hat\gamma^{45})^\ast\ .
\end{equation}
So $C\otimes\hat{C}\epsilon_+^\ast$ has the opposite sign in its $\hat\gamma^{45}$
eigenvalue to $\epsilon_+$, making it possible to identify it as $\epsilon_-$.
To conclude, the spinors $\epsilon$ satisfying (\ref{killing-5d}) can be regarded as
forming a set of $8$ Poincare SUSY Q and $8$ conformal SUSY S in 6d perspective, which
closes into itself under Hermitian conjugation.
These $8$ complex or $16$ real Killing spinors will be the SUSY of our 5d SYM.

As a more general twisting, one can choose different $SO(2)$ embeddings in $SO(5)$,
which generically result in a 5d theory with $8$ preserved SUSY upon circle reduction. One
introduces the twisting which covariantizes
\begin{equation}
  \nabla_\tau\ \rightarrow\ \ \nabla_\tau+\frac{i}{2r}\left(\Delta\hat\gamma^{45}
  +(1-\Delta)\hat\gamma^{12}\right)
\end{equation}
on spinors, where $\Delta$ is a real constant. By following the discussions
of the last paragraph, one finds that the reduced 5d theory preserves $8$ SUSY,
which satisfies $i\hat\gamma^{45}=i\hat\gamma^{12}=\pm 1$ projection for $\epsilon_\pm$,
respectively.

Now let us capture some key aspects of the 5d Abelian gauge theory obtained by reducing
the 6d free tensor theory on the circle, with the above R-symmetry twist. In the
$r\rightarrow\infty$ limit, we simply get the maximal SYM in 5 dimension. The coupling
to the background curvature yields various mass terms in the Abelian theory. From the
viewpoint of the 6d theory on $S^5\times S^1$, the mass terms come from
two sources. Firstly, when one radially quantizes the 6d theory, all 5 real scalars
acquire the conformal mass terms with mass $m=\frac{2}{r}$, since the free scalars have
dimension $2$. This yields the 6d mass terms
\begin{equation}
  \frac{2}{r^2}(\phi^a)^2+\frac{2}{r^2}(\phi^i)^2
\end{equation}
with $a=1,2,3$, $i=4,5$, in the convention that the kinetic terms are
$\frac{1}{2}(\partial\phi^a)^2+\frac{1}{2}(\partial\phi^i)^2$. In 5d, extra
contributions to the mass terms are induced from the kinetic term with $\tau$
derivatives, since we now have the $SO(2)$ twists. There is no extra contribution
for $\phi^a$, but the $\tau$ derivatives on $\phi^i$ and the fermions $\lambda$ are
twisted as
\begin{eqnarray}
  \nabla_\tau\phi^i&\rightarrow&\nabla_\tau\phi^i-\frac{i}{r}\epsilon^{ij}\phi^j\nonumber\\
  \nabla_\tau\lambda&\rightarrow&(\nabla_\tau-\frac{i}{2r}\hat\gamma^{45})\lambda\ .
\end{eqnarray}
respectively. The 6d kinetic terms thus provide extra contribution to the 5d masses
\begin{equation}
  \frac{1}{2}(\nabla_\tau\phi^i)^2+\frac{1}{2}\lambda^\dag\nabla_\tau\lambda
  +\frac{2}{r^2}(\phi^a)^2+\frac{2}{r^2}(\phi^i)^2\rightarrow\frac{2}{r^2}(\phi^a)^2
  +\frac{3}{2r^2}(\phi^i)^2-\frac{i}{4r}\lambda^\dag\hat\gamma^{45}\lambda\ .
\end{equation}
Adding the last scalar and fermion mass terms to be maximal SYM action (with obvious
covariantization with the 5-sphere metric), one is supposed to obtain an Abelian action
which preserves $16$ SUSY. We shall explicitly show that the theory preserves $16$ SUSY
with above masses in section 2.2, with a non-Abelian completion.

The case with general $SO(2)$ embedding can be studied as well. The resulting scalar
and fermion mass terms are given by
\begin{equation}
  \frac{4-(1-\Delta)^2}{2r^2}(\phi^a)^2
  +\frac{4-\Delta^2}{2r^2}(\phi^i)^2-\frac{i}{4r}\lambda^\dag
  \left(\Delta\hat\gamma^{45}+(1-\Delta)\hat\gamma^{12}\right)\lambda\ .
\end{equation}
We shall come back to this version of non-Abelian theory with $8$ SUSY later.

Before proceeding, we illustrate the nature of the 6d partition functions that we expect
our 5d calculations to capture, with the example of 6d Abelian $(2,0)$ theory on
$S^5\times\mathbb{R}$.
Up to global rotations and charge conjugation, the BPS bound given by a chosen pair
of $Q$ and $S$ via $\{Q,S\}$ in 6d is given by
\begin{equation}\label{bound}
  \epsilon\geq 2(R_1+R_2)+j_1+j_2+j_3\ ,
\end{equation}
where $R_1$ is the $SO(2)$ R-symmetry we used to twist the time derivative.
$R_2$ is another Cartan of $SO(5)$ in the orthogonal 2-plane basis, and
$j_1,j_2,j_3$ are three $SO(6)$ Cartans, again in the three orthogonal 2-plane basis.
The twist above with $8$ SUSY uses $\Delta R_1\!+\!(1-\Delta)R_2$. There is one
Poincare supercharge $Q$ saturating the above energy bound, which has $R_1=R_2=\frac{1}{2}$,
$j_1=j_2=j_3=-\frac{1}{2}$, $\epsilon=\frac{1}{2}$.
The index which counts BPS states saturating this bound is studied
in \cite{Kinney:2005ej,Bhattacharya:2008zy}. It is defined as
\begin{equation}\label{superconformal}
  {\rm Tr}\left[(-1)^Fe^{-\beta^\prime\{Q,S\}}x^{3\epsilon+j_1+j_2+j_3}
  y^{R_1-R_2}a^{j_1}b^{j_2}c^{j_3}\right]
\end{equation}
with a constraint $abc=1$. $\beta^\prime$ is the usual regulator in the Witten index.
For the $U(1)$ $(2,0)$ theory,
the full index $Z$ is given by the Plethystic (or multi-particle)
exponential of the letter index $z$ \cite{Bhattacharya:2008zy}
\begin{equation}\label{abelian-index}
  z=\frac{x^6(y+y^{-1})-x^8(ab+bc+ca)+x^{12}}{(1-x^4a)(1-x^4b)(1-x^4c)}\ ,\ \
  Z=x^{\epsilon_0}\exp\left[\sum_{n=1}^\infty\frac{1}{n}z(x^n,y^n,a^n,b^n,c^n)\right]\ .
\end{equation}
$\epsilon_0$ is the `index version' of the vacuum Casimir energy of the Abelian theory
on $S^5\times\mathbb{R}$. See appendix B. The terms in the numerators can be easily
understood from the BPS fields in the free Abelian tensor multiplet. The first two
terms come from two complex scalars (among $5$ real) taking charges
$\Phi^{(R_1,R_2)}_{(j_1,j_2,j_3)}=\Phi^{(1,0)}_{(0,0,0)}$ and $\Phi^{(0,1)}_{(0,0,0)}$.
The next $3$ terms come from three chiral fermions with charges
$\Psi^{(R_1,R_2)}_{(j_1,j_2,j_3)}=\Psi^{(+,+)}_{(-,+,+)}$, $\Psi^{(+,+)}_{(+,-,+)}$
and $\Psi^{(+,+)}_{(+,+,-)}$, where $\pm$ denote $\pm\frac{1}{2}$. The final term $+x^{12}$
is for a fermionic constraint coming from a component of the Dirac equation which
contains BPS fields and derivatives only,
$(\slash\hspace{-.25cm}\partial\Psi)^{(+,+)}_{(+,+,+)}=0$. The three factors in the
denominator come from acting three holomorphic derivatives to the above BPS fields
and constraints, which have $R_1=R_2=0$ and $(j_1,j_2,j_3)=(1,0,0)$, $(0,1,0)$ and
$(0,0,1)$.

The contribution
$x^{\epsilon_0}$ is normally ignored in the literature on the superconformal index, but
should be there as an overall multiplicative factor in path integral approaches
\cite{Aharony:2003sx}. Of course
the index (\ref{superconformal}) can be defined in non-Abelian theories
with the same $(2,0)$ superconformal algebra.

There are two interesting limits of this general index which we consider in this paper.
Firstly, one can take $x\rightarrow 0$, $y\rightarrow\infty$, keeping $x^6y\equiv q$ fixed.
The letter index $z$ becomes $z=q$ in this limit, yielding
\begin{equation}
  Z=\lim_{x\rightarrow 0}(x^{\epsilon_0})\frac{1}{1-q}\ .
\end{equation}
The first factor either goes to zero or infinity. As we explain in appendix B,
the Casimir energy for the 6d $(2,0)$ theory is expected to be negative. In any case,
one normally considers the remaining factor $\frac{1}{1-q}$,
which is the half-BPS partition function which acquires contribution from operators made
of a single complex scalar. Its non-Abelian version for $U(N)$ gauge
group \cite{Bhattacharyya:2007sa} is explained in section 3.3.

Another limit, which is of more interest to us in this paper, is obtained by taking
all but one fugacity variables to be $1$, so that more cancelations are expected to
appear than the general superconformal index. We call this the unrefined index.
To explain this limit, we start by noting that
the supercharges chosen above commutes with $\epsilon-R_1$. The fugacity conjugate to
this charge is a particular combination of the four fugacities $x,y,a,b(,c)$.
We turn off three fugacities to $1$ apart from the one conjugate to $\epsilon-R_1$,
which we call $q$. More concretely, we first rewrite the measure in (\ref{superconformal})
using the BPS relation $\epsilon=2R_1+2R_2+j_1+j_2+j_3$:
\begin{equation}
  x^{3\epsilon+j_1+j_2+j_3}y^{R_1-R_2}a^{j_1}b^{j_2}c^{j_3}=
  x^{4\epsilon}(yx^{-2})^{R_1}(yx^2)^{-R_2}a^{j_1}b^{j_2}c^{j_3}\ .
\end{equation}
Then setting $a=b=c=x^2y=1$, and defining $q\equiv x^4$, the measure becomes
$q^{\epsilon-R_1}$. Note that, as
the half-BPS energy bound in 6d is $\epsilon\geq 2|R_1|$, $\epsilon-R_1$ is positive
definite for all states. Rewriting the unrefined letter index (\ref{abelian-index})
using $q$ only, one obtains
\begin{equation}
  z=\frac{q+q^2-3q^2+q^3}{(1-q)^3}=\frac{q}{1-q}\ ,\ \
  Z=q^{\epsilon_0}PE\left[\frac{q}{1-q}\right]=q^{\epsilon_0}\prod_{n=1}^\infty
  \frac{1}{1-q^n}\ .
\end{equation}
Although the second limit is very different from the first limit above for the half-BPS
states, it has a special property associated with the same $16$ SUSY. Namely,
$\epsilon-R_1$ commutes with exactly the same $16$ supercharges preserved by
the half-BPS states considered in the last paragraph. Superconformal indices can be
defined by choosing any $2$ mutually conjugate supercharges $Q$, $S$ among them.
One would obtain the same result no matter which pair one chooses.

The R-symmetry twist we introduced above for the Abelian theory provides the
chemical potential to $R_1$ as well, so that we weight the states by
$e^{-\beta(\epsilon-R_1)}$. The $16$ SUSY of the 5d theory refers to those in 6d
which commutes with $\epsilon-R_1$. Thus, we expect the partition function of
this 5d theory with $16$ SUSY to be the second limit of the superconformal index,
with identification $q=e^{-\beta}$ of the fugacity and the gauge coupling.
During detailed calculations in later sections, we shall
use localization by picking any of the $16$ SUSY of the theory. The result is
guaranteed to be the same from 5d perspective as the path integral preserves all
$16$ SUSY, among which we only use a pair. This is consistent with our observation
in the previous paragraph from the 6d perspective, that same result will be obtained
no matter what supercharges one chooses to define the index.

An important property of the second limit is that the information on the vacuum
Casimir energy is not lost. So if one can compute the partition function for
non-Abelian theories, the $N^3$ scaling is supposed to be calculable
in a microscopic way.

The information on the above two limiting cases of the superconformal index is
all encoded in the following simplified index. Namely, we consider an unrefined
index which contains only two chemical potentials conjugate to $\epsilon-R_1$,
$\epsilon-R_2$. In (\ref{abelian-index}), this amounts to turning off $a,b,c$ and
keeping $x$, $y$ only. We weight the states as $q_1^{\epsilon-R_1}q_2^{\epsilon-R_2}$.
The resulting letter index for the Abelian theory becomes
\begin{equation}
  z=\frac{q_1q_2^2+q_1^2q_2-3q_1^2q_2^2+q_1^3q_2^3}{(1-q_1q_2)^3}\ .
\end{equation}
The first term in the numerator comes from a complex scalar which defines the half-BPS
states. The scaling limit $q_1\rightarrow 0$, $q_2\rightarrow\infty$ which keeps
$q\equiv q_1q_2^2$ finite takes the above letter index to $q$, which yields the desired
half-BPS partition function for the Abelian theory. In the 5d reduction, the parameters $\beta,\Delta$ are related to
$q_1$, $q_2$ by
\begin{equation}\label{two-chemical}
  q_1=e^{-\beta\Delta}\ ,\ \ q_2=e^{-\beta(1-\Delta)}\ ,\ \
  q\equiv q_1q_2^2=e^{-\beta(2-\Delta)}\ .
\end{equation}
The half-BPS limit amounts to taking
\begin{equation}
  \beta\rightarrow\infty\ ,\ \ \Delta\rightarrow 2\ ,\ \ \beta(2-\Delta)={\rm fixed}\ .
\end{equation}
In section 3.2, we shall explain the structure of the $S^5$ partition with two parameters
$\beta,\Delta$, which is supposed to capture the 6d index
${\rm Tr}[(-1)^Fq_1^{\epsilon-R_1}q_2^{\epsilon-R_2}]$.

With more twists with the global symmetries of the theory, including R-symmetries
above as well as spatial rotations, it will be possible to obtain a 5d action
which preserves less supersymmetries, and presumably on a squashed $S^5$.
Then one can reduce the Abelian theory along the circle to obtain a 5d theory, and
calculate the partition function after a non-Abelian generalization which can be used
to study the general superconformal index \cite{Kinney:2005ej,Bhattacharya:2008zy} of
the 6d $(2,0)$ theory. This problem  is studied in our later work \cite{Kim:2012qf}.

\subsection{Non-Abelian theories}

We generalize the above Abelian 5d theory on the 5-sphere, with
$SO(3)\times SO(2)$ subgroup of $SO(5)$ R-symmetry preserved by the curvature
coupling, to the non-Abelian gauge groups. We find that the action is
\begin{eqnarray}\label{action}
  S&=&\frac{1}{g_{YM}^2}\int d^5x\sqrt{g}\
  {\rm tr}\left[\frac{1}{4}F_{\mu\nu}F^{\mu\nu}+\frac{1}{2}D_\mu\phi^I D^\mu\phi^I
  +\frac{i}{2}\lambda^\dag\gamma^\mu D_\mu\lambda-\frac{1}{4}[\phi^I,\phi^J]^2
  -\frac{i}{2}\lambda^\dag\hat\gamma^I[\lambda,\phi^I]\right.\nonumber\\
  &&\hspace{2.7cm}\left.+\frac{4}{2r^2}(\phi^a)^2+\frac{3}{2r^2}(\phi^i)^2
  -\frac{i}{4r}\lambda^\dag\hat\gamma^{45}\lambda-\frac{1}{3r}\epsilon_{abc}\phi^a
  [\phi^b,\phi^c]\right]\ ,
\end{eqnarray}
where $I,J=1,2,3,4,5$, $a\!=\!1,2,3$, $i\!=\!4,5$ are the vector indices of $SO(5)$
R-symmetry. $\gamma^\mu$ and $\hat\gamma^I$ are $4\times 4$
gamma matrices for the spatial/internal $SO(5)$, respectively.
This action is invariant under the following $16$ supersymmetries:
\begin{eqnarray}\label{SUSY}
  -i\delta A_\mu&=&\frac{i}{2}\lambda^\dag\gamma_\mu\epsilon-\frac{i}{2}\epsilon^\dag
  \gamma_\mu\lambda\\
  -i\delta\phi^I&=&-\frac{1}{2}\lambda^\dag\hat\gamma^I\epsilon+\frac{1}{2}\epsilon^\dag
  \hat\gamma^I\lambda\nonumber\\
  -i\delta\lambda&=&\frac{1}{2}F_{\mu\nu}\gamma^{\mu\nu}\epsilon+iD_\mu\phi^I\gamma^\mu
  \hat\gamma^I\epsilon-\frac{i}{2}[\phi^I,\phi^J]\hat\gamma^{IJ}\epsilon
  +\frac{2i}{r}\phi^a\hat\gamma^{a45}\epsilon+\frac{i}{r}\phi^i\hat\gamma^{i}\hat\gamma^{45}
  \epsilon\nonumber\\
  -i\delta\lambda^\dag&=&-\frac{1}{2}\epsilon^\dag\gamma^{\mu\nu}F_{\mu\nu}+
  i\epsilon^\dag\hat\gamma^I\gamma^\mu D_\mu\phi^I-\frac{2i}{r}\epsilon^\dag\hat\gamma^{45a}\phi^a-\frac{i}{r}\epsilon^\dag
  \hat\gamma^{45}\hat\gamma^i\phi^i+\frac{i}{2}\epsilon^\dag\hat\gamma^{IJ}[\phi^I,\phi^J]
  \nonumber
\end{eqnarray}
where $\epsilon$ satisfies
\begin{equation}
  \nabla_\mu\epsilon=\frac{1}{2r}\gamma_\mu\hat\gamma^{45}\epsilon\ ,\ \
  \nabla_\mu\epsilon^\dag=-\frac{1}{2r}\epsilon^\dag\gamma_\mu\hat\gamma^{45}
\end{equation}
on $S^5$. As we already explained with the Abelian theories, we take $\epsilon_+$ with
$i\hat\gamma^{45}=+1$ eigenvalues, which is related to $\epsilon_-$ with $-1$
eigenvalue by a symplectic charge conjugation.

We explain the reality property of the action and SUSY transformation in some detail.
Imposing the symplectic Majorana conditions for all matter and Killing spinors, the
action (\ref{action}) is real apart from the last term which is cubic in the scalars.
Also, we note that the SUSY transformations between scalars-fermions are all real,
while those between vector-fermions are all imaginary, i.e. violating reality condition.
The factor $-i$ we inserted on the left hand sides of (\ref{SUSY}) guarantees the
above property.\footnote{Compared to the 5d maximal SYM action on the flat Euclidean space,
perhaps this $-i$ factor is unconventional. In the last case, the reality condition is
often ignored as we are in a Euclidean space.}
So in the path integral with this action, the $16$ SUSY transformations should be
regarded as symmetry transformations associated with changes of some integration
contours. The localization method that we shall use later in this paper applies with
such a complexification.

Technically, we started with the Abelian theory on $S^5$ obtained by a
Scherk-Schwarz reduction from 6d, and then added non-Abelian terms to SUSY and action,
trying to secure $16$ SUSY. We think the complex transformation and action are compulsory
consequences of this analysis, as we also tried but failed to find other real versions.
At least one can motivate why gauge fields-fermion part of the transformation could be
imaginary from the Abelian theory (in which case the action is actually real).
Consider some part of $16$ SUSY, e.g. $8$ SUSY that we consider in the later part
of this section. This choice of $8$ SUSY provides a notion of vector and hypermultiplets.
The supersymmetric reduction of the free hypermultiplet part is quite clear, and we
find no reason to ruin the reality of the SUSY transformation in this part. However,
the gauge field/fermion part seems somewhat subtle. In the Lorentzian theory on
$S^5\times\mathbb{R}$, the self-dual 3-form condition $H_{\mu\nu\rho}=\frac{1}{2}\epsilon_{\mu\nu\rho\alpha\beta}H^{\alpha\beta 0}$ can be
solved by naturally taking $F_{\mu\nu}=H_{\mu\nu 0}$ to be independent momentum-like fields,
subject to 6d Bianchi identity for $H_{MNP}$. In the Euclidean theory on
$S^5\times\mathbb{R}$,
covariant self-dual condition cannot be imposed. Still we want to secure the number
of degrees of freedom as this will be natural for getting the correct physics.
If we stick to the definition of $F_{\mu\nu}$ as $H_{\mu\nu 0}$, one would have to
continue $F_{\mu\nu}$ to $H_{\mu\nu 6}=-iH_{\mu\nu 0}=-iF_{\mu\nu}$ along
$\tau=it$. This extra factor of $i$ would make the vector-fermion
SUSY transformation to be imaginary. Combined with the formal SUSY checks that we did,
which independently yielded imaginary transformations, we feel that (\ref{SUSY}) is
somewhat inevitable.\footnote{However, one could have imposed different reality conditions
on various fields. For instance, the choice of \cite{Hosomichi:2012ek} is
different from ours in many places. Although not all the prescriptions in
\cite{Hosomichi:2012ek} are well motivated to us, by suitable analytic continuations
or complexifications we can make half of our SUSY to fit into theirs.}

One can check that the supersymmetry algebra is $SU(4|2)$. Firstly, one can obtain
the following commutation relations
\begin{eqnarray}\label{superalgebra}
  [\delta_1,\delta_2]\phi^{a} &=& 2i\epsilon_1^\dagger\gamma^\mu\epsilon_2 D_\mu\phi^a+2i\epsilon_1^\dagger\hat\gamma^J\epsilon_2[\phi^J,\phi^a]
  +\frac{4i}{r}\epsilon_1^\dagger\hat\gamma^{ab}\hat\gamma^{45}\epsilon_2\phi^b  \nonumber\\
  &=& L_v \phi^a +i[\Lambda,\phi^a]+\frac{2i}{r}\epsilon^{abc}\epsilon^\dagger_1
  \hat\gamma^b\epsilon_2 \phi^c \,, \\
  \left. \right.[\delta_1,\delta_2]\phi^{i} &=&2i\epsilon_1^\dagger\gamma^\mu\epsilon_2 D_\mu\phi^i+2i\epsilon_1^\dagger\hat\gamma^J\epsilon_2[\phi^J,\phi^i]
  -\frac{2i}{r}\epsilon_1^\dagger\epsilon_2\epsilon^{ij}\phi^j  \nonumber \\
  &=& L_v \phi^i +i[\Lambda,\phi^i]+\frac{i}{r}\epsilon_1^\dagger\epsilon_2
  \epsilon^{ij}\phi^j\,, \nonumber\\
  \left. \right. [\delta_1,\delta_2]A_\mu &=& 2i\epsilon_1^\dagger\gamma^\nu\epsilon_2F_{\nu\mu} +2\epsilon_1^\dagger\hat\gamma^I\epsilon_2D_\mu\phi^I-\frac{2}{r}\epsilon^{ij}
  \epsilon_1^\dagger\gamma_\mu\hat\gamma^i\epsilon_2\phi^j \nonumber\\
  &=& L_v A_\mu +D_\nu\Lambda\nonumber\\
  \left.\right.[\delta_1,\delta_2]\lambda&=&L_v\lambda+i[\Lambda,\lambda]
  +\frac{1}{4}\Theta^{\mu\nu}\gamma_{\mu\nu}\lambda-i\epsilon^\dag\epsilon_2
  \hat\gamma^{45}\lambda-2i\epsilon_1^\dag\hat\gamma^a\epsilon_2\hat\gamma^{a45}\lambda
  +({\rm eqn\ of\ motion})\nonumber
\end{eqnarray}
where
\begin{eqnarray}
  &&v^\mu = 2i\epsilon^\dagger_1\gamma^\mu\epsilon_2 \ ,\ \
  \Lambda =-2i \epsilon^\dagger_1\gamma^\mu\epsilon_2 A_\mu + 2\epsilon^\dagger_1\tilde\gamma^I\epsilon_2\phi^I \,, \nonumber\\
  &&L_v\phi^i= v^\mu\partial_\mu \phi^i \ ,\ \ L_v\phi^a=v^\mu\partial_\mu \phi^a \,,
  L_v A_\mu = v^\nu\partial_\nu A_\mu + \partial_\mu v^\nu A_\nu\ ,\nonumber\\
  &&\Theta^{\mu\nu}=\nabla^{[\mu}\xi^{\nu]}+\xi^\lambda\omega_\lambda^{\ \mu\nu}\ .
\end{eqnarray}
In 6d $SU(4|2)$, the bosonic
subgroup is $SU(4)\times SU(2)\times U(1)$, where the $U(1)$ part is $\epsilon-R_1$.
By dimensional reduction to $S^5$, one is only left
with $-R_1$ which appears on the right hand side of (\ref{superalgebra}) as rotations
by $\epsilon^{ij}\phi^j$. Also, using the following Fierz identities
\begin{eqnarray}
 \hspace*{-1cm}(\epsilon_1^\dag\gamma_\nu\epsilon_2)
 (\epsilon_3^\dag\gamma^{\mu\nu}\hat\gamma^{45}\epsilon_4)
 \!\!&\!=\!&\!\!-\frac{1}{4}(\epsilon_1^\dag\epsilon_4)
 (\epsilon_3^\dag\gamma^{\mu\nu}\gamma_\nu\hat\gamma^{45}\epsilon_2)
 -\frac{1}{4}(\epsilon_1^\dag\gamma^\alpha\epsilon_4)
 (\epsilon_3^\dag\gamma^{\mu\nu}\gamma_\alpha\gamma_\nu\hat\gamma^{45}\epsilon_2)+\frac{1}{8}
 (\epsilon_1^\dag\gamma^{\alpha\beta}\epsilon_4)(\epsilon_3^\dag\gamma^{\mu\nu}
 \gamma_{\alpha\beta}\gamma_\nu\hat\gamma^{45}\epsilon_2)\nonumber\\
 \hspace*{-1cm}(\epsilon_1^\dag\gamma^{\mu\nu}\hat\gamma^{45}\epsilon_2)
 (\epsilon_3^\dag\gamma_\nu\epsilon_4)\!\!&\!=\!&\!\!-\frac{1}{4}(\epsilon_1^\dag\epsilon_4)
 (\epsilon_3^\dag\gamma_\nu\gamma^{\mu\nu}\hat\gamma^{45}\epsilon_2)
 -\frac{1}{4}(\epsilon_1^\dag\gamma^\alpha\epsilon_4)
 (\epsilon_3^\dag\gamma_\nu\gamma_\alpha\gamma^{\mu\nu}\hat\gamma^{45}\epsilon_2)+\frac{1}{8}
 (\epsilon_1^\dag\gamma^{\alpha\beta}\epsilon_4)(\epsilon_3^\dag\gamma_\nu
 \gamma_{\alpha\beta}\gamma^{\mu\nu}\hat\gamma^{45}\epsilon_2)\nonumber
\end{eqnarray}
and taking all spinors to belong to $\epsilon_-$, one can check for
$v^\mu=2i\epsilon_1\gamma^\mu\epsilon_2$, $w^\mu=2i\epsilon_3^\dag\gamma^\mu\epsilon_4$
that
\begin{eqnarray}
  [v,w]^\mu&=&\mathcal{L}_vw^\mu=-\frac{4}{r}(\epsilon_1^\dag\gamma_\nu\epsilon_2)
  \left(\epsilon_3^\dag\gamma^{\mu\nu}\hat\gamma^{45}\epsilon_4\right)
  +\frac{4}{r}\left(\epsilon_1^\dag\gamma^{\mu\nu}\hat\gamma^{45}\epsilon_2\right)
  (\epsilon_3^\dag\gamma_\nu\epsilon_4)\nonumber\\
  &=&\frac{8}{r}(\epsilon_1^\dag\epsilon_4)(\epsilon_3^\dag\gamma^\mu\hat\gamma^{45}\epsilon_2)
  -\frac{8}{r}(\epsilon_1^\dag\gamma^\mu\epsilon_4)(\epsilon_3^\dag\hat\gamma^{45}\epsilon_2)\ .
\end{eqnarray}
Normalizing spinors as $\epsilon_\alpha^\dag\epsilon_\beta=\delta_{\bar\alpha\beta}$
where $\alpha,\beta=1,2,3,4$ are for ${\bf 4}$ of $SO(6)$, one obtains
\begin{equation}
  [v_{\bar{\alpha}\beta},v_{\bar{\gamma}\delta}]^\mu=-\frac{4}{r}
  \left(\delta_{\beta\bar{\gamma}}v_{\bar{\alpha}\delta}^\mu
  -\delta_{\bar{\alpha}\delta}v_{\bar{\gamma}\beta}^\mu\right)\ ,
\end{equation}
which is forming the desired $SU(4)\sim SO(6)$ algebra. $SU(2)$ part of the algebra
is also easily visible as rotations on $\phi^a$. So we interpret it as the 5d reduction
of $SU(4|2)\subset OSp(8|4)$ superconformal group for the 6d $(2,0)$ theory, commuting
with $\epsilon-R_1$.

By taking all $\epsilon_i$'s to be $\epsilon_-$ above, we obtained the
anti-commutation relations of the type $\{Q,S\}$. The commutation relations of
the form $\{Q,Q\}$ or its conjugate $\{S,S\}$ can be studied by taking $\epsilon_1$
to belong to $\epsilon_-$ and $\epsilon_2$ to belong to $\epsilon_+$ in
(\ref{superalgebra}). Then, one finds
\begin{equation}
  \epsilon_1^\dag\gamma^\mu\epsilon_2=0\ ,\ \
  \epsilon_1^\dag\hat\gamma^{a}\epsilon_2=0\ ,\ \ \epsilon_1^\dag\epsilon_2=0
\end{equation}
by studying $i\hat\gamma^{45}=(i\hat\gamma^{45})^\dag$ eigenvalues. Thus, the
bosonic elements of the superalgebra do not extend beyond $SU(4|2)$. For instance,
the analysis for 5d SCFT with $F(4)$ symmetry would have yielded $\{Q,Q\}\sim P$,
$\{S,S\}\sim K$ as in \cite{Kim:2012gu}, but they naturally do not appear in our case.

In the next section, we shall use the localization method to perform the path integral
for the partition function. To this end, we attempt to make some part of the supersymmetry
algebra to hold off-shell. The most important requirement is that the single supercharge,
or a pair of conjugate supercharges, which we choose to perform localization calculation
takes the required algebra (nilpotency) off-shell. We take $8$ of our $16$ SUSY, and
decompose the field into the vector and hypermultiplets. The vector multiplet part of the
algebra is made off-shell for all $8$ SUSY by introducing $3$ auxiliary fields, while hypermultiplet part of the algebra is made off-shell only for a subset which includes
a pair of Hermitian SUSY generators. This strategy is all spelled out in
\cite{Hosomichi:2012ek}.

With the internal gamma matrices chosen as (\ref{internal-gamma}), the $8$ SUSY
are chosen by taking $\hat\gamma^3\epsilon=-\epsilon$. The internal charge conjugation
matrix is taken to be $\hat{C}=\hat\gamma^{25}=i\sigma^2\otimes\sigma^3$.
One can write the $8$ SUSY and $16$ component fermion $\lambda$ as
\begin{equation}
  \epsilon=\left(\begin{array}{c}\epsilon^1\\ \epsilon^2\end{array}\right)\otimes
  \left(\begin{array}{c}1\\0\end{array}\right)\ ,\ \ \lambda=
  \left(\begin{array}{c}\chi^1\\ \chi^2\end{array}\right)\otimes
  \left(\begin{array}{c}1\\0\end{array}\right)+
  \left(\begin{array}{c}\psi^1\\ \psi^2\end{array}\right)\otimes
  \left(\begin{array}{c}0\\1\end{array}\right)\ .
\end{equation}
$\epsilon_A,\chi_A,\psi_A$ for $A=1,2$ can be regarded as $SU(2)$ spinors. This $SU(2)$
symmetry is broken in the action by curvature couplings, and only the Cartan generator
proportional to $\sigma^3$ is a symmetry. The $SO(5)$ origin of this $U(1)$ can be easily
traced by noticing $\hat\gamma^{12}=i\sigma^3\otimes{\bf 1}_2$,
$\hat\gamma^{45}=i\sigma^3\otimes\sigma^3$. The $U(1)$ acts on $\chi_A$ as a simultaneous
rotation on $12$ and $45$ planes, while on $\psi_A$ as opposite rotation on the two 2-planes.
For later use, we take a complex 4-component spinor $\psi$ on $S^5$ as $\psi\equiv\psi^2$.
The first component $\psi^1$ is related to $\psi$ by a symplectic-Majorana conjugation using
$SO(5)\times SU(2)$, inherited from our $SO(5)\times SO(5)$ symplectic-Majorana conjugation.
Let us also define the scalars as
\begin{equation}
  \phi\equiv\phi^3\ ,\ \ q^1\equiv\frac{1}{\sqrt{2}}(\phi^4-i\phi^5)\ ,\ \
  q^2\equiv\frac{1}{\sqrt{2}}(\phi^1+i\phi^2)\ .
\end{equation}
The real scalar $\phi$ participates in the vector multiplet, while $q^A$
belong to the adjoint hypermultiplet.

For the vector multiplet, we introduce three auxiliary fields $D^I$, whose
on-shell values become
\begin{equation}
  D^I=-(\sigma^I)^A_{\ B}[q^B,\bar{q}_A]-\frac{i}{r}\delta^I_3\phi\ .
\end{equation}
The off-shell Lagrangian that we shall write in a moment is invariant under
\begin{eqnarray}
  -i\delta A_\mu&=&i\chi^\dag\gamma_\mu\epsilon\\
  -i\delta\phi&=&\chi^\dag\epsilon\nonumber\\
  -i\delta\chi&=&\frac{1}{2}F_{\mu\nu}\gamma^{\mu\nu}\epsilon-iD_\mu\phi\gamma^\mu\epsilon
  +\frac{1}{r}\phi\sigma^3\epsilon+iD^I\sigma^I\epsilon\nonumber\\
  -i\delta\chi^\dag&=&-\frac{1}{2}F_{\mu\nu}\epsilon^\dag\gamma^{\mu\nu}
  -i\epsilon^\dag\gamma^\mu D_\mu\phi-\frac{1}{r}\epsilon^\dag\sigma^3\phi
  -i\epsilon^\dag\sigma^ID^I\nonumber\\
  -i\delta D^I&=&D_\mu\chi^\dag\gamma^\mu\sigma^I\epsilon-[\phi,\chi^\dag]\sigma^I\epsilon
  -\frac{i}{2r}\chi^\dag\sigma^I\sigma^3\epsilon\ .
\end{eqnarray}
The off-shell SUSY algebra including the $8$ Killing spinors is $SU(4|1)$, and is given by
\be
\left[\delta_1,\delta_2\right]\!A_\mu &=& \xi^\nu\partial_\nu A_\mu +\partial_\mu\xi^\nu A_\nu + D_\mu \Lambda , \nn \\
\left[\delta_1,\delta_2\right]\!\phi^3 &=& \xi^\mu\partial_\mu\phi^3 +i[\Lambda,\phi^3]  + \rho\phi^3,\nn \\
\left[\delta_1,\delta_2\right]\!\chi &=& \xi^\mu \partial_\mu\chi +\frac{1}{4}\Theta_{\mu\nu}\gamma^{\mu\nu}\lambda+i[\Lambda,\chi] +\frac{3}{2}\rho\chi +\frac{3}{4}R^{IJ}\sigma^{IJ}\chi,\nn\\
\left[\delta_1,\delta_2\right]\!{\rm D}^I &=& \xi^\mu\partial_\mu {\rm D}^I + i[\Lambda,{\rm D}^I] +2\rho {\rm D}^I+3R^{IJ}{\rm D}^J\ ,
\ee
where
\be
&&\xi^\mu = 2i\bar\epsilon_1\gamma^\mu\epsilon_2\ ,\ \
\Lambda = -2i\bar\epsilon_1\gamma^\mu\epsilon_2A_\mu+2\bar\epsilon_1\epsilon_2\phi^3, \nn \\
&&\Theta^{\mu\nu} = D^{[\mu}\xi^{\nu]} +\xi^\lambda\omega_\lambda^{\mu\nu},\ ,\ \
\rho = \frac{2i}{5}D_\mu(\bar\epsilon_1\gamma^\mu\epsilon_2), \nn \\
&&R^{IJ} = \frac{2i}{5}(\bar\epsilon_1\gamma^\mu\sigma^{IJ}D_\mu\epsilon_2-D_\mu\bar\epsilon_1\gamma^\mu \sigma^{IJ}\epsilon_2)\ .
\ee
These results are all found in \cite{Hosomichi:2012ek}.

We also consider an off-shell generalization of the hypermultiplet algebra.
As the off-shell generalization of the whole $8$ SUSY algebra cannot be achieved
with a finite number of auxiliary fields, we follow the strategy of \cite{Hosomichi:2012ek}
and demand that we have a single off-shell nilpotent supercharge, with which
one can do localization calculations. In other words, we are interested in a SUSY which
satisfies $\delta^2=0$ off-shell (up to a bosonic symmetry generator) with a given
commuting spinor $\epsilon$ parameter.
With a bosonic $\epsilon$ chosen among the $8$ SUSY generators explained above,
we follow \cite{Hosomichi:2012ek} and consider another bosonic spinor parameter
$\hat\epsilon$ satisfying
\begin{equation}
  \epsilon^\dag\epsilon=\hat\epsilon^\dag\hat\epsilon\ ,\ \
  (\epsilon^A)^TC\hat\epsilon^{B^\prime}=0\ ,\ \ \epsilon^\dag\gamma^\mu\epsilon
  +\hat\epsilon^\dag\gamma^\mu\hat\epsilon=0\ .
\end{equation}
One introduces two auxiliary complex fields $F^A$, having $0$ on-shell values,
and consider the following SUSY transformation with a commuting Killing spinor
(which reduces to our on-shell SUSY upon taking $F^A=0$):
\be\label{hyper-off-shell}
    \delta q^A &=& \sqrt{2}i(\epsilon^\dagger)^A\psi \ , \quad \delta \bar{q}_A =-\sqrt{2}i\psi^\dagger\epsilon_A \nn \\
    \delta \psi &=& \sqrt{2}\left[-D_\mu q_A\gamma^\mu \epsilon^A  +[\phi^3,q_A]\epsilon^A- \frac{3i}{2r}q_A(\sigma^3)^A_{\ \ B}\epsilon^B-\frac{i}{2r}q_A\epsilon^A - iF_{A'}\hat{\epsilon}^{A'}\right] \nn \\
    \delta \psi^\dagger &=& \sqrt{2}\left[\epsilon^\dagger_A\gamma^\mu D_\mu \bar{q}^A  +\epsilon^\dagger_A[\bar{q}^A,\phi^3] -i \frac{3}{2r}\epsilon^\dagger_A(\sigma^3)^A_{\ \ B}\bar{q}^B-\frac{i}{2r}\epsilon^\dagger_A\bar{q}^A - i(\hat{\epsilon}^\dagger)_{A'}\bar{F}^{A'}\right] \nn \\
    \delta F^{A'} &=& \sqrt{2}(\hat{\epsilon}^\dagger)^{A'}\left[-\gamma^\mu D_\mu\psi + \frac{i}{2r}\psi -[\phi^3,\psi]-\sqrt{2}[\chi_A, q^A]\right] \nn \\
    \delta\bar{F}_{A'} &=& \sqrt{2}\left[-D_\mu\psi^\dagger\gamma^\mu -\frac{i}{2r}\psi^\dagger+[\psi^\dagger,\phi^3] -\sqrt{2}[\bar{q}_A,(\chi^\dagger)^A]\right]\hat\epsilon_{A'}\ .
\ee
This is a special case of \cite{Hosomichi:2012ek} which has $-\frac{1}{2r}q_A\epsilon^A$,
$\frac{1}{2r}\psi$ terms on the right hand sides with a choice of their mass parameters.
The SUSY algebra for a given commuting $\epsilon$ is
\be
    \delta^2q^A &=&\xi^\mu\partial_\mu q^A + i [\Lambda, q^A] + \frac{3}{4} R^{IJ} (\sigma^{IJ} q )^A + \frac{1}{2r}q^A \nn \\
    \delta^2\psi &=& \xi^\mu \partial_\mu \psi +\frac{1}{4}\Theta_{\mu\nu}\gamma^{\mu\nu}\psi+i\Lambda\psi + \frac{1}{2r}\psi \nn \\
    \delta^2F^{A'} &=& \xi^\mu \partial_\mu F^{A'} + i[\Lambda,F^{A'}] +\frac{5}{4}\hat{R}^{IJ}(\hat\sigma^{IJ}F)^{A'} +\frac{1}{2r} F^{A'}\ ,
\ee
where
\be
    &&\xi^\mu = -i\epsilon^\dagger\gamma^\mu\epsilon \ , \ \
    \Lambda = i\epsilon^\dagger\gamma^\mu\epsilon A_\mu+\phi \ , \ \
    \Theta^{\mu\nu} = D^{[\mu}\xi^{\nu]} +\xi^\lambda\omega_\lambda^{\mu\nu} \ , \nn \\
    &&R^{IJ} =- \frac{2i}{5}\epsilon^\dagger\sigma^{IJ}\gamma^\mu D_\mu\epsilon \ , \ \
    \hat{R}^{IJ}=\frac{2i}{5}\hat\epsilon^\dagger\hat\sigma^{IJ}\gamma^\mu D_\mu\hat\epsilon\ .
\ee
In the above off-shell formulation, the Lagrangian
invariant under the above $8$ SUSY transformations is given by
\begin{eqnarray}\label{off-shell-action}
  \mathcal{L}&=&\frac{1}{g_{YM}^2}{\rm tr}\left[\frac{1}{4}F_{\mu\nu}F^{\mu\nu}
  +\frac{1}{2}(D_\mu\phi)^2+\frac{i}{2}\chi^\dag\gamma^\mu D_\mu\chi
  -\frac{1}{2}D^ID^I-\frac{i}{r}D^3\phi+\frac{5}{2r^2}\phi^2
  -\frac{i}{2}\chi^\dag[\phi,\chi]+\frac{1}{4r}\chi^\dag\sigma^3\chi\right.\nonumber\\
  &&+|D_\mu q^A|^2+i\psi^\dag\gamma^\mu D_\mu\psi+|[\phi,q^A]|^2
  -D^I(\sigma^I)^A_{\ B}[q^B,\bar{q}_A]-F^{A^\prime}\bar{F}_{A^\prime}
  -\frac{i}{r}\phi[q^A,\bar{q}_A]+\frac{3}{r^2}|q^1|^2+\frac{4}{r^2}|q^2|^2\nonumber\\
  &&\left.+i\psi^\dag[\phi,\psi]
  +\sqrt{2}i\psi^\dag[\chi_A,q^A]-\sqrt{2}i[\bar{q}_A,\chi^{\dag A}]\psi+\frac{1}{2r}\psi^\dag\psi\right]
\end{eqnarray}
The integration contours for $D^I$, ${\rm Re}(F^A)$, ${\rm Im}(F^A)$ are taken to be
on the imaginary axes.

We can generalize the theory preserving 8 SUSY with a continuous parameter $\Delta$,
whose Abelian version we introduced in section 2.1 (corresponding to a generalized
Scherk-Schwarz reduction). The generalized Lagrangian is
\be
    \mathcal{L}_{YM} &=& \frac{1}{g_{YM}^2}{\rm tr}\Big[\frac{1}{4}F_{\mu\nu}F^{\mu\nu}+\frac{1}{2}(D_\mu\phi^3)^2 + |D_\mu q^A|^2
    +\frac{5}{2r^2}(\phi^3)^2 +\frac{15}{4r^2}|q^A|^2 -\frac{1}{2}{\rm D}^I {\rm D}^I-\frac{i}{r}\phi^3 {\rm D}^3\nn \\
    &&+\left([\bar{q}_A,\phi^3]+i\frac{1-2\Delta}{2r}\bar{q}_A\right)\left([\phi^3,q^A]+i\frac{1-2\Delta}{2r}q^A\right)-\bar{q}_A(\sigma^I)^A_{\ \ B}\left([{\rm D}^I,q^B]-\delta^I_3\frac{1-2\Delta}{2r^2}q^B\right) \nn \\
    &&+\frac{i}{2}\chi^\dagger \gamma^\mu D_\mu\chi +i\psi^\dagger \gamma^\mu D_\mu \psi + \frac{1}{4r}\chi^\dagger\sigma^3\chi -\bar{F}_{A'}F^{A'}\nn \\
     &&- \frac{i}{2}\chi^\dagger[\phi^3,\chi]+i\psi^\dagger\left([\phi^3,\psi]+i\frac{1-2\Delta}{2r}\psi\right) +\sqrt{2}i\psi^\dagger[\chi_A,q^A] -\sqrt{2}i[\bar{q}_A,\chi^\dagger]\psi\Big]\ .
\ee
When $\Delta =1$, it becomes our previous action with $16$ SUSY.
It is invariant under
\be\label{hyper-off-shell}
    \delta q^A &=& \sqrt{2}i(\epsilon^\dagger)^A\psi \ , \quad \delta \bar{q}_A =-\sqrt{2}i\psi^\dagger\epsilon_A \nn \\
    \delta \psi &=& \sqrt{2}\left[-D_\mu q_A\gamma^\mu \epsilon^A+[\phi^3,q_A]\epsilon^A- \frac{3i}{2r}q_A(\sigma^3)^A_{\ \ B}\epsilon^B +i\frac{1-2\Delta}{2r}q_A\epsilon^A- iF_{A'}\hat{\epsilon}^{A'}\right] \nn \\
    \delta \psi^\dagger &=& \sqrt{2}\left[\epsilon^\dagger_A\gamma^\mu D_\mu \bar{q}^A  +\epsilon^\dagger_A[\bar{q}^A,\phi^3] - \frac{3i}{2r}\epsilon^\dagger_A(\sigma^3)^A_{\ \ B}\bar{q}^B +i\frac{1-2\Delta}{2r}\epsilon^\dagger_A\bar{q}^A- i(\hat{\epsilon}^\dagger)_{A'}\bar{F}^{A'}\right] \nn \\
    \delta F^{A'} &=& \sqrt{2}(\hat{\epsilon}^\dagger)^{A'}\left[-\gamma^\mu D_\mu\psi  -i\frac{1-2\Delta}{2r}\psi -[\phi^3,\psi]+\sqrt{2}i[\chi_A, q^A]\right] \nn \\
    \delta\bar{F}_{A'} &=& -\sqrt{2}\left[D_\mu\psi^\dagger\gamma^\mu-i\frac{1-2\Delta}{2r}\psi^\dagger-[\psi^\dagger,\phi^3] -\sqrt{2}i[\bar{q}_A,(\chi^\dagger)^A]\right]\hat\epsilon_{A'}
\ee
and same SUSY transformation on vector multiplet fields.
One can identify the fields and parameters in our theory and \cite{Hosomichi:2012ek} as
\begin{equation}
    \phi^3 =- i\sigma_{HST} \ , \quad \chi = -i\lambda_{HST} \ ,
    \quad i \sigma^I D^I = D_{HST} \ ,
    q^A = q^A_{HST} \ , \quad \psi = \sqrt{2}\psi_{HST}\ .
\end{equation}
Our parameter $\Delta-\frac{1}{2}$ is proportional to their hypermultiplet mass
associated with
a global symmetry. The off-shell SUSY algebra for the vector multiplet is the same, while
the off-shell algebra for a given commuting Killing spinor for hypermultiplet becomes
\be
    \delta^2q^A &=&\xi^\mu\partial_\mu q^A + i [\Lambda, q^A] + \frac{3}{4} R^{IJ}
    (\sigma^{IJ} q )^A - \frac{1-2\Delta}{2r}q^A \nn \\
    \delta^2\psi &=& \xi^\mu \partial_\mu \psi +\frac{1}{4}\Theta_{\mu\nu}\gamma^{\mu\nu}\psi+i\Lambda\psi - \frac{1-2\Delta}{2r}\psi \nn \\
    \delta^2F^{A'} &=& \xi^\mu \partial_\mu F^{A'} + i[\Lambda,F^{A'}] +\frac{5}{4}\hat{R}^{IJ}(\hat\sigma^{IJ}F)^{A'} -\frac{1-2\Delta}{2r} F^{A'}\ .
\ee
In section 3.3, we shall use this theory to calculate the perturbative partition function,
which we suggest would be part of a more general superconformal index.

\section{5-sphere partition function as a 6d index}

In this section, we study the partition function of the maximal SYM on $S^5$
and the theory with $8$ SUSY that we considered in the previous section.

We first consider the theory with $16$ SUSY. We choose a commuting Killing spinor
$\epsilon$ to be a linear combination $\epsilon=\epsilon^++\epsilon^-$, where
$\epsilon^\pm$ satisfy the following projection conditions
\begin{equation}
  \sigma^3\epsilon^\pm=\pm\epsilon^\pm\ ,\ \
  \gamma^5\epsilon^\pm=\mp i\gamma^{12}\epsilon^\pm=\pm i\gamma^{34}\epsilon^\pm=\epsilon^\pm\ .
\end{equation}
The explicit expressions for $\epsilon^\pm$ are (see appendix A, $\eta_\pm$ there)
\begin{equation}
  \epsilon^\pm=e^{\pm\frac{3i}{2}y}\epsilon^\pm_0\ ,
\end{equation}
where constant spinors $\epsilon^\pm_0$ are conjugate to each other as
$(\epsilon^+_0)^\ast=C\otimes(i\sigma^2)\epsilon^-_0$. $y$ is the angle coordinate
of the Hopf fiber of $S^5$, over a $\mathbb{CP}^2$ base. The following spinor bilinears
will be useful:
\begin{equation}
  v^\mu=\epsilon^\dag\gamma^\mu\epsilon\ ,\ \ J_{\mu\nu}=\nabla_\mu v_\nu
  =-2i\bar\epsilon^+\gamma_{\mu\nu}\epsilon^+\ (=e^1\wedge e^2-e^3\wedge e^4)\ .
\end{equation}
$J_{\mu\nu}$ is the Kahler 2-form of $\mathbb{CP}^2$, and $v^\mu$ is the translation
generator along the fiber $y$ direction. They satisfy $\nabla_\rho J_{\mu\nu}=2v_{[\mu}g_{\nu]\rho}$.
With this $\epsilon$, we can add any term to the Lagrangian $\mathcal{Q}V$ which is exact in the corresponding supercharge $\mathcal{Q}$, without changing the value of the final integral. This
property relies on the property that the chosen $\mathcal{Q}$ is nilpotent, $\mathcal{Q}^2=0$.
Actually, since the chosen Killing spinor $\mathcal{Q}$ is real, it amounts to picking one Poincare
supercharge $Q$ with its conjugate conformal supercharge $S$, and taking a real linear combination
of the two. Thus one actually finds
\begin{equation}
  \mathcal{Q}^2\sim\{Q,S\}=({\rm symmetry\ generator})\ ,
\end{equation}
where the right hand side comes from a suitable
combination of the bosonic generators appearing in the $\{Q,S\}$ part of the $SU(4|2)$
algebra. Thus, only when we choose $V$ in the $\mathcal{Q}$-exact deformation $\mathcal{Q}V$
to be neutral under the rotation of $\{Q,S\}$ (which we will do), one is guaranteed
not to change the partition function by deformation.

$\mathcal{Q}$-exact deformations that we introduce are
\begin{eqnarray}\label{exact-vector}
  \delta\left((\delta\chi)^\dag\chi\right)&=&\frac{1}{2}F_{\mu\nu}F^{\mu\nu}
  -\frac{1}{4}\epsilon^{\mu\nu\rho\sigma\tau}v_\mu F_{\nu\rho}F_{\sigma\tau}
  +(D_\mu\phi)^2+\left(\frac{1}{r}\phi+iD^3\right)^2-(D^1)^2-(D^2)^2\nonumber\\
  &&-i\chi^\dag\gamma^\mu D_\mu\chi-i[\phi,\chi^\dag]\chi+\frac{1}{r}\chi^\dag
  \sigma^3\chi-\frac{1}{2r}\chi^\dag v_\mu\gamma^\mu\sigma^3\chi
  -\frac{i}{4r}J_{\mu\nu}\chi^\dag\gamma^{\mu\nu}\chi
\end{eqnarray}
for the vector multiplet, and
\be\label{exact-hyper}
    &&\frac{1}{2}\delta\Big((\delta\psi)^\dagger\psi+\psi^\dagger(\delta\psi^\dagger)^\dagger\Big)
    \nonumber\\
    &&= |D_\mu q^A|^2-\frac{i}{r}v^\mu\bar{q}\sigma^3 D_\mu q -\frac{i}{r}v^\mu \bar{q}D_\mu q
    +\frac{1}{r^2}\bar{q}_1 q^1 +\frac{4}{r^2}\bar{q}_2 q^2 +|[\phi^3,q^A]|^2 - \bar{F}_{A'}F^{A'}\nonumber\\
    &&\hspace{0.5cm}+i\psi^\dagger\gamma^\mu D_\mu \psi -\frac{1}{2r}v^\mu \psi^\dagger\gamma_\mu\psi-\frac{i}{4r}J^{\mu\nu}\psi^\dagger\gamma_{\mu\nu}\psi +
    i\psi^\dagger[\phi^3,\psi]
\ee
for the hypermultiplet. Here, the commuting Killing spinors are normalized to satisfy
$\epsilon^\dag\epsilon=1$, and traces are assumed for every terms.\footnote{We add
two conjugate terms to form $V$ in the hypermultiplet part (\ref{exact-hyper}),
as this simplifies the determinant calculation significantly.}
It is easy to see that the corresponding $V$'s that we introduced above all
commute with $\{Q,S\}$. As $V$ are chosen to take the form of $(\delta\Phi)^\dag\Phi$
for various fields $\Phi$,
the charge of $V$ under $\{Q,S\}$ is basically the inverse of the charge carried by
the chosen SUSY generator $\delta$. As this is a linear combination of $Q,S$, it suffices
to show that  $Q,S$ are both neutral under the rotation of $\{Q,S\}$. This trivially follows
from the following Jacobi identities (with $\{Q,Q\}=\{S,S\}=0$)
\begin{equation}
  [\{Q,S\},Q]=0\ ,\ \ [\{Q,S\},S]=0\ .
\end{equation}
Thus we are allowed to introduce the above $\mathcal{Q}$-exact deformations.

\subsection{Perturbative partition function and Casimir energies}

Turning on the above $\mathcal{Q}$-exact deformations and taking their
coefficients to be large, one is led to a Gaussian path integral around a set
of saddle points satisfying
\begin{equation}\label{saddle}
  F_{\mu\nu}=\frac{1}{2}\sqrt{g}\epsilon_{\mu\nu\alpha\beta\gamma}
  v^\alpha F^{\beta\gamma}\ ,\ \ D_\mu\phi=0\ ,\ \ D^3=\frac{i}{r}\phi\ ,\ \
  D^1=D^2=0\ ,\ \ q_1=q_2=0\ ,\ \ F^{1^\prime}=F^{2^\prime}=0\ ,
\end{equation}
while taking all fermion fields to zero. These equations can be easily obtained
by studying the vanishing SUSY condition, or alternatively by taking the bosonic part of
the $\mathcal{Q}$-exact deformations (\ref{exact-vector}), (\ref{exact-hyper}) to be zero.
See also \cite{Kallen:2012cs,Hosomichi:2012ek} which study the same equations.

The first equation of (\ref{saddle}) is for the self-dual Yang-Mills instantons
on the $\mathbb{CP}^2$ base (in the convention that the Kahler 2-form of $\mathbb{CP}^2$
is anti-self-dual), while any component of the gauge field along the Hopf fiber
is demanded to be zero from $v^\mu F_{\mu\nu}=0$. The configurations solving this
equation are called `contact instantons' in some literatures, and recently studied
on general contact
manifolds, including $S^5$ \cite{Harland:2011zs,Wolf:2012gz}. In particular,
\cite{Wolf:2012gz} explores the twistor construction of this equation, which
could probably be used to get a better understanding of its solutions.
If the topological quantum number for these instantons on $\mathbb{CP}^2$ is nonzero,
one would get various non-perturbative corrections to the partition function. We shall
study them in the next subsection, and focus on the perturbative part here.

With $F_{\mu\nu}=0$, one can take the gauge connection to zero on $S^5$.
The only nonzero fields at the saddle point are $D^3$
and $\phi$ satisfying $D^3=\frac{i}{r}\phi$, where $\phi$ is a constant
Hermitian matrix. The saddle point is thus parameterized by the Hermitian matrix $\phi$,
which we should exactly integrate over after all other Gaussian
fluctuations are integrated out. The integration over $\phi$ will come with various factors
of integrands. Part of them will come from the contributions from the determinants of
quadratic fluctuations, which we shall turn to in a while. There is also a factor of integrand
that one obtains by plugging in the saddle point values of the fields into the original action.
Plugging in nonzero $\phi$ and $D^3$ into (\ref{off-shell-action}), this becomes
\begin{equation}
  e^{-S_0}\ ,\ \ S_0=\frac{1}{g_{YM}^2}\int d^5x\sqrt{g}\frac{4}{r^2}{\rm tr}\phi^2=
  \frac{4\pi^3 r^3}{g_{YM}^2}{\rm tr}\phi^2=\frac{2{\rm tr}(\pi r\phi)^2}{\beta}
  \equiv\frac{2\pi^2{\rm tr}\lambda^2}{\beta}\ ,
\end{equation}
where $\int d^5x\sqrt{g}=\pi^3r^5$ on a 5-sphere with radius $r$,
$\frac{4\pi^2}{g_{YM}^2}=\frac{1}{r_1}=\frac{2\pi}{r\beta}$ yields
$\frac{4\pi^3 r^3}{g_{YM}^2}=\frac{2\pi^2r^2}{\beta}$, and we defined
$\lambda\equiv r\phi_0$ at the last step. The natural justification of
the $g_{YM}$ vs. $\beta$ relation we use here is given in section 3.2.

From the vector multiplet bosons, one has to diagonalize the differential operator
appearing in the following quadratic fluctuations in the $\mathcal{Q}$-exact deformation
($\phi$ fluctuations simply decouple to yield a constant factor, which cancels out
with other constant factors):
\be\label{vector-eqn}
    &&\frac{1}{2}F_{\mu\nu}F^{\mu\nu} -\frac{1}{4} \epsilon^{\mu\nu\lambda\rho\sigma}v_\mu F_{\nu\lambda}F_{\rho\sigma} \nn \\
    &=& A^\mu\left(-D^2 \delta_\mu^\nu + D_\mu D^\nu + 4\delta_\mu^\nu -2(J_{\mu\lambda}v\cdot D + 2v_{[\mu}J_{\lambda]\rho}D^\rho) g^{\lambda\nu}\right)A_\nu\ .
\ee
Using the basis of the vector spherical harmonics introduced in appendix A to diagonalize
the differential operator, one obtains the following determinant:
\begin{eqnarray}
  {\det}_{V,b}&=&\prod_{\alpha\in{\rm root}}
  \prod_{k=0}^\infty\left(k+4+ir\alpha(\phi)\right)^{\frac{(k+1)(k+2)^2(k+3)}{12}}
  (k+ir\alpha(\phi))^{\frac{(k+1)(k+2)^2(k+3)}{12}-2\times\frac{(k+1)(k+2)}{2}}\nonumber\\
  &&\times\prod_{k=1}^\infty\prod_{m=-k+1}^k\left(k^2+4k-2m+9+r^2\alpha(\phi)^2
  \right)^{\frac{(k+2)((k+2)^2-m^2)}{8}}\ .
\end{eqnarray}
See appendix A for the derivation.
From the vector multiplet fermions, one obtains
\begin{eqnarray}
  \hspace*{-1cm}{\det}_{V,f}&=&\prod_{\alpha\in{\rm root}}
  \prod_{k=0}^\infty(k+4+ir\alpha(\phi))^{\frac{(k+1)(k+2)^2(k+3)}{12}}
  (k+ir\alpha(\phi))^{\frac{(k+1)(k+2)^2(k+3)}{12}-\frac{(k+1)(k+2)}{2}}
  (k+3+ir\alpha(\phi))^{\frac{(k+1)(k+2)}{2}}\nonumber\\
  &&\times\prod_{k=1}^\infty\prod_{m=-k+1}^k
  \left(k^2+4k-2m+9+r^2\alpha(\phi)^2\right)^{\frac{(k+2)((k+2)^2-m^2)}{8}}\ .
\end{eqnarray}
Dividing the two contributions, one obtains
\begin{equation}
  \frac{{\det}_{V,f}}{{\det}_{V,b}}=\prod_{\alpha\in{\rm root}}
  \prod_{k=0}^\infty(k+3+ir\alpha(\phi))^{\frac{(k+1)(k+2)}{2}}
  \prod_{k=1}^\infty(k+ir\alpha(\phi))^{\frac{(k+1)(k+2)}{2}}
  =\prod_{\alpha\in{\rm root}}\prod_{k=1}^\infty(k+ir\alpha(\phi))^{k^2+2}\ .
\end{equation}
This agrees with the result found in \cite{Kallen:2012cs}.

From the hypermultiplet, one obtains from the two complex scalars $q_1$, $q_2$ the following:
\begin{equation}
  {\rm det}_{H,b}=\prod_{\alpha\in{\rm root}}
  \prod_{k=0}^\infty\frac{1}{\left((k+2)^2+r^2\alpha(\phi)^2\right)^{\frac{(k+1)(k+2)^2(k+3)}{12}}}
  \prod_{m=-k}^k\frac{1}{\left(k^2+4k+1+2m+r^2\mathcal{\mathcal{\mathcal{}}}\alpha(\phi)^2\right)^{\frac{(k+2)((k+2)^2-m^2)}{8}}}
\end{equation}
where $m=k,k-2,k-4,\cdots,-k$. From hypermultiplet fermions,
\begin{eqnarray}
  {\rm det}_{H,f}&=&\prod_{\alpha\in{\rm root}}
  \prod_{k=1}^\infty(k+2+ir\alpha(\phi))^{\frac{(k+1)(k+2)^2(k+3)}{6}
  -\frac{(k+1)(k+2)}{2}}\\
  &&\times\prod_{k=0}^\infty(k+1+ir\alpha(\phi))^{\frac{(k+1)(k+2)}{2}}\prod_{m=-k+1}^k
  \left(k^2+4k+1+2m+r^2\alpha(\phi)^2\right)^{\frac{(k+2)((k+2)^2-m^2)}{8}}\ .\nonumber
\end{eqnarray}
The net hypermultiplet determinant is
\begin{equation}
  \frac{{\rm det}_{H,f}}{{\rm det}_{H,b}}=\prod_{\alpha\in{\rm root}}
  \prod_{k=1}^\infty\frac{1}{(k+ir\alpha(\phi))^{k^2}}\ .
\end{equation}
Again see appendix A for the derivation.

Combining the contributions from vector and hypermultiplets,
one obtains the following perturbative determinant
\begin{equation}\label{final-measure}
  \prod_{\alpha\in{\rm root}}\prod_{k=1}^\infty(k+ir\alpha(\phi))^2
  =\prod_{\alpha\in{\rm root}}\prod_{k=1}^\infty(k^2+r^2\alpha(\phi)^2)=
  \prod_{\alpha\in{\rm root}}\frac{2\pi\sinh(\pi r\alpha(\phi))}{\pi r\alpha(\phi)}\ .
\end{equation}
Here, we used $\prod_{k=1}^\infty k^2\!=\!2\pi$ after zeta function regularization
\cite{Kallen:2012cs}. The integration over the Hermitian matrix can be replaced
by an integration over the eigenvalues with
the Vandermonde measure inserted, which cancels $\alpha(\phi)$ in the
denominator of (\ref{final-measure}). Combining it with the classical Gaussian
measure, and defining dimensionless variables $\lambda=r\phi$, one obtains
\begin{equation}
  Z_{\rm pert}=\frac{1}{|W|}\int d\lambda\
  e^{-\frac{2\pi^2{\rm tr}(\lambda^2)}{\beta}}
  \prod_{\alpha\in{\rm root}}2\sinh(\pi\alpha(\lambda))\ ,
\end{equation}
where $W$ is the Weyl group. One thus finds that the perturbative part of
the partition function, with $16$ SUSY, takes the form of the pure Chern-Simons
partition function on $S^3$ \cite{Marino:2002fk}. See also
\cite{Kapustin:2009kz,Tierz:2002jj} for some later studies of the same expression.

For simplicity, let us first consider the case with $U(N)$ gauge group in detail.
Pure Chern-Simons partition function with $U(N)$ gauge group is
\cite{Marino:2002fk,Kapustin:2009kz}
\begin{eqnarray}\label{U(N)-CS}
  Z_{CS}&=&\frac{1}{N!}\int\prod_id\lambda_ie^{-ik\pi\lambda_i^2}
  \prod_{i\neq j}2\sinh\left(\pi\lambda_{ij}\right)\\
  &=&\nonumber\frac{(-1)^{\frac{N(N-1)}{2}}e^{-\pi iN^2/4}e^{-\frac{\pi i}{6k}N(N^2-1)}}
  {k^{N/2}}\prod_{m=1}^{N-1}\left[2\sin\frac{\pi m}{k}\right]^{N-m}\ .
\end{eqnarray}
Comparing with our partition function, one should replace
$-\frac{i\pi}{k}$ by $\frac{\beta}{2}$. Thus one finds
\begin{eqnarray}\label{perturbative}
  Z_{\rm pert}&=&(-1)^{N(N-1)/2}\left(\frac{i\beta}{2\pi}\right)^{N/2}
  e^{-\pi iN^2/4}e^{\frac{N(N^2-1)}{12}\beta}
  \prod_{m=1}^{N-1}\left[i(e^{\frac{m\beta}{2}}-e^{-\frac{m\beta}{2}})\right]^{N-m}\\
  &=&(-1)^{N(N-1)/2}\left(\frac{i\beta}{2\pi}\right)^{N/2}
  e^{-\pi iN^2/4}i^{N(N-1)/2}
  e^{\frac{N(N^2-1)}{6}\beta}\prod_{m=1}^{N-1}(1-e^{-\beta m})^{N-m}\nonumber
\end{eqnarray}
where we used $\sum_mm(N-m)=\frac{N(N^2-1)}{6}$. The factors of $i$'s combine to
be $i^{N^2/2}e^{-\pi i N^2/4}=1$, and we shall not be careful about the
possible overall minus sign. Thus, regarding $q\equiv e^{-\beta}$ as the fugacity
of $\epsilon-R_1$ in the 6d theory, the perturbative contribution itself would have
taken the form of an index, supposing that we can somehow trade away the prefactor
$\left(\frac{\beta}{2\pi}\right)^{N/2}$. We shall see in the next subsection that,
combining this factor with the non-perturbative contribution will make the latter
to be an index. So we ignore this piece in this subsection and proceed.

More generally, for the gauge group $G$ with rank $r$, the 3-sphere Chern-Simons
partition function is given by \cite{Witten:1988hf,Marino:2011nm}
\begin{equation}\label{general-CS}
  Z_{CS}=[\det(C)]^{1/2}\frac{i^{\frac{|G|}{2}-r}}{k^{r/2}}e^{-\frac{\pi i}{6k}c_2|G|}
  \prod_{\alpha>0}2\sin\frac{\pi(\alpha\cdot\rho)}{k}\ ,
\end{equation}
where $|G|$ is the dimension of the gauge group, $c_2$ is the dual Coxeter number,
and $C$ is the inverse matrix of the inner product in the weight space (or Cartan
matrix for simply connected gauge group $G$). 
$\rho$ is the Weyl vector which is the summation of all fundamental weights.

To get the correct information on the BPS state degeneracies, we will also have to
include the non-perturbative corrections, which we discuss in the next subsection.
However, from (\ref{perturbative}) one immediately observes a multiplicative factor
$e^{-\beta(\epsilon_0)_{\rm pert}}$ with
\begin{equation}
  (\epsilon_0)_{\rm pert}=-\frac{N(N^2-1)}{6}
\end{equation}
for $U(N)$. For general gauge group, one finds from (\ref{general-CS}) that
$(\epsilon_0)_{\rm pert}$ becomes
\begin{equation}\label{pert-casimir}
  (\epsilon_0)_{\rm pert}=-\frac{c_2|G|}{6}\ ,
\end{equation}
where $c_2$ is the dual Coxeter number and $|G|$ is the dimension of the
semi-simple part of the gauge group $G$.
See the next subsection for a nonperturbative correction to this result (subleading in
the large $N$ limit). This factor can naturally be interpreted as the `vacuum
energy' or the Casimir energy. However, one should be careful about the identification of
$\epsilon_0$ as the Casimir energy, as one has to pick a regularization when one
computes the vacuum energy. For instance, in free QFT, the Casimir energy is the summation of
all bosonic mode frequencies minus the fermionic mode frequencies, divided by $2$. In a radially
quantized CFT, one can employ the zeta function regularization or the energy
regularization as done, e.g. in \cite{Aharony:2003sx}. However, our result
above can be regarded as a `Casimir energy'
obtained by using $\epsilon-R_1$ as a regulator, as this is the only charge which can appear
in this index. In many theories, including 4d SCFTs admitting free theory limits, we illustrate
that different regularizations lead to different $\epsilon_0$. However, we observe that the
index Casimir energy contains useful information on the degrees of freedom of the theory.
In particular, in all 4d SCFT examples that we study in appendix B, we find that the index
Casimir energy is always proportional to the Casimir energy by a universal coefficient, and
is also a particular linear combination of the $a$ and $c$ central charge of the CFT.
Thus, we think our index Casimir energy could also be an interesting measure of the
degrees of freedom.

Coming back to our case, the coefficient in front of $c_2|G|$ has no reason to agree
with the true Casimir energy, due to the usage of an index version of regularization
and renormalization. Indeed, the calculation of the large $N$ Casimir energy of
$AdS_7\times S^4$ from gravity yields \cite{Awad:2000aj}
\begin{equation}
  \epsilon_0=-\frac{5N^3}{24\ell}\ ,
\end{equation}
where $\ell$ is the $AdS_7$ radius. The coefficients $-\frac{1}{6}$ and $-\frac{5}{24}$ in
front of $N^3$ are indeed different. However, our $\epsilon_0$ robustly reproduces the expected $N^3$ behavior in the large $N$ limit, which we regard as a significant microscopic
evidence supporting that $N$ M5-branes exhibit $N^3$ some of degrees. It should be interesting to study the gravity dual of (\ref{pert-casimir}). It is also curious
that the finite rank index Casimir energy from the perturbative part is proportional to
$c_2|G|$, which is the anomaly coefficient of the $ADE$ $(2,0)$ theory in 6d
\cite{Harvey:1998bx}. See, however, section 3.3 for a subleading correction that is
contained in a non-perturbative correction that we propose.

It should be very desirable to pursue the virtue of the index Casimir energy that we
have studied here (and in appendix B), and try to relate it to other measures of degrees
of freedom such as central charges, as we illustrate in appendix B with concrete examples
in 4d.

To better understand the perturbative expansion structure of $Z_{\rm pert}$,
We expand it in the large $N$ limit with small 't Hooft coupling, $\beta\rightarrow 0$,
$N\rightarrow\infty$, $N\beta={\rm fixed}\ll 1$. The perturbative `free energy'
$F_{\rm pert}=-\log Z_{\rm pert}$ is expanded as
\begin{eqnarray}\label{weak-coupling}
  F_{\rm pert}&=&-\frac{N}{2}\log\frac{\beta}{2\pi}-\frac{\beta N(N^2-1)}{6}
  -\sum_{n=1}^N(N-n)\log(1-e^{-n\beta})\nonumber\\
  &\rightarrow&-\frac{N^2}{2}\log(N\beta)+\frac{3N^2}{4}
  +N^2\sum_{n=1}^\infty a_n(N\beta)^n
\end{eqnarray}
with some $\mathcal{O}(1)$ coefficients $a_n$, where we used
\begin{eqnarray}
  \sum_{n=1}^N n\log n&=&\frac{N^2}{2}\log N-\frac{N^2}{4}+\frac{N}{2}\log N
  +\frac{1}{12}\log N+\mathcal{O}(1)\nonumber\\
  \log N!&=&N\log N-N+\frac{1}{2}\log(2\pi N)+\mathcal{O}(N^{-1})
\end{eqnarray}
to obtain the first two leading terms in $N\beta$.
Here, at each order in $N\beta$, we only showed
the leading terms in $N$. Especially, the last infinite sum is acquiring contributions
from the planar diagrams. Naturally, the leading term in the weak coupling expansion
scales like $N^2$. It is also of some interest to study a sub-leading term
at the 2-loop ($\sim N^3\beta$) order, to study the 5d aspect of the 6d Casimir
energy that we obtained above. From the exact expression given on the first
line of (\ref{weak-coupling}), this order term comes from two sources. It first comes
from the second term $\beta(\epsilon_0)_{\rm pert}=-\frac{\beta N(N^2-1)}{6}$.
Also, the last summation which takes the form
$-\sum_n d_n\log(1-e^{-\beta E_n})$ yields a term at the same order,
$\frac{\beta}{2}\sum_nd_n E_n$ with $d_n=N-n$ and $E_n=n$.
Adding them, one obtains the following net (finite $N$) 2-loop contribution
\begin{equation}\label{2-loop-combination}
  -\frac{\beta N(N^2-1)}{6}+\frac{\beta}{2}\sum_{n=1}^Nn(N-n)
  =-\frac{\beta N(N^2-1)}{12}\ .
\end{equation}
So in the weak coupling regime, the information on the Casimir energy
$(\epsilon_0)_{\rm pert}$ in $F_{\rm pert}$ totally goes to the 2-loop order,
but also became ambiguous at this order by combining with an extra
contribution from $\beta/2\sum_n d_n E_n$.

We also work out a strong coupling large $N$ limit of $F_{\rm pert}$,
keeping $\beta$ finite ($N\beta\rightarrow\infty$). It turns out that
the leading behavior is the same as the 't Hooft large $N$ limit with
$\lambda={\rm fixed}\gg 1$, although the sub-leading terms are differently
organized in the two limits. The former limit is perhaps
more interesting, as this regime admits a dual gravity description in a
Euclidean $AdS_7$ which is supersymmetrically compactified along the time
direction with finite radius. On the first line of (\ref{weak-coupling}),
the second term is dominant in this strong coupling large $N$ limit:
\begin{equation}\label{strong}
  F_{\rm pert}\sim-\frac{\beta N^3}{6}\ .
\end{equation}
So it acquires contribution only from the large $N$ Casimir energy.
Even with the instanton correction provided in the next subsection (proved in
\cite{Kim:2012qf}), this is the dominant term in the full free energy.

\begin{figure}[t!]
  \begin{center}
    \includegraphics[width=12cm]{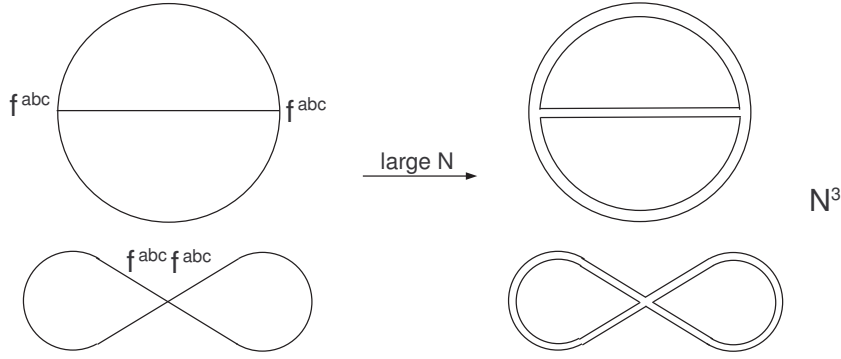}
\caption{2-loop diagrams and the large $N$ double-line diagrams with $N^3$ scalings
for $SU(N)$}\label{2-loop}
  \end{center}
\end{figure}
With the above understandings, it is easy to trace how the $N^3$ scaling, or more precisely
the $c_2|G|$ factor, appears in the Casimir energy, from the viewpoint of perturbative QFT.
$\beta\epsilon_0$ appears in $F_{\rm pert}$ at the sub-leading 2-loop order $\beta\sim
g_{YM}^2$ at weak coupling. Considering possible 2-loop vacuum bubbles such as those shown
as the Feynman diagrams of Fig \ref{2-loop}, it is clear that the group theoretic factors
are always $f^{abc}f^{abc}=c_2|G|$. Also, in the large $N$ double line notation for $U(N)$,
the appearance of 3 single loops naturally yields the $N^3$ scaling. Strictly speaking,
this argument does not say that $(\epsilon_0)_{\rm pert}$ itself shows the
$N^3$ behavior, but just that the combination $(\epsilon_0)_{\rm pert}+\frac{1}{2}\sum_nd_nE_n$
in (\ref{2-loop-combination}) does. But also with our exact result (\ref{perturbative}),
it still looks like a heuristic 5d insight of the appearance of $N^3$ in
$\epsilon_0$.\footnote{This was our original motivation that the $N^3$ scaling
in $\epsilon_0$ could appear from 5d gauge theories.}

Actually, such a group theoretic argument at $\mathcal{O}(g_{YM}^2)$ applies to
any quantum field theories with adjoint fields, in any dimension. For instance,
this is basically the reason why $N(N^2-1)$ or $c_2|G|$ appears in the pure Chern-Simons
partition functions (\ref{U(N)-CS}), (\ref{general-CS}). However, for generic
adjoint QFT's, this is no more than the standard 't Hooft planar contribution at
a particular sub-leading order, or a group theory of quadratic Casimir $f^{abc}f^{abc}$
at finite $N$. It is only because we have a higher dimensional
interpretation (with $g_{YM}^2$ being related to the inverse temperature or the
6th direction's radius in our case) that we can take this $c_2|G|$ or $N^3$
scaling as the physics of 6d $(2,0)$ theory. Also, for generic adjoint QFT's, there
is no guarantee that the strong coupling large $N$ limit would be anything like
(\ref{strong}).\footnote{For some special QFT's, like
pure Chern-Simons theory on $S^3$ whose partition function takes the same form as
our $Z_{\rm pert}$ with $\beta\sim i/k$, one might be able to say more on this
term which scales like $N^3$ (still subleading at weak 't Hooft coupling).
We are not sure if this has any meaning at all, perhaps in a different
physical context.}

We also note that, from the viewpoint of our 5-sphere partition function, it is not
clear at this stage whether $ADE$ gauge theories have any special status to have 6d UV
fixed points, as many arguments go similarly for other gauge groups $BCFG$. For instance,
the index nature of the Chern-Simons index could appear from (\ref{general-CS}), from the
expansion of the sine factors. Just like the $U(N)$ case that we explained, there
is $\beta^{r/2}$ prefactor and possibly some non-integral constant factor which will
obstruct $Z_{\rm pert}$ from being an index. Like the $U(N)$ case, all such factors
should combine with the non-perturbative part to be an index, for the $S^5$ partition
function to be interpretable as a 6d index. It could be that this non-perturbative
corrections, combined with the above prefactors, may violate the 6d index
structure for non-$ADE$ gauge groups. However, one should not confuse the 6d gauge
group and 5d gauge group which appears after compactification. For instance, suitably
twisted compactifications of 6d $ADE$ theories can yield all $BCFG$ gauge groups in 5d
\cite{Tachikawa:2011ch}. Our $BCFG$ partition functions could thus be `twisted indices,'
similar to \cite{Zwiebel:2011wa}.

%

\subsection{Nonperturbative corrections and $AdS_7$ gravity duals}

To motivate the studies on possible non-perturbative corrections to our partition
function, let us first go back to the 6d index explained in section 2.1, and
study it for the free Abelian 6d theory. This free theory index would be also important
in the $U(N)$ theories, as the overall $U(1)$ degrees are decoupled from the rest
which forms the interacting $A_n$ type $(2,0)$ theory.
In section 2.1, the `letter index' was shown to be $z=\frac{q}{1-q}$, and
the full index is given by
\begin{equation}\label{free-index}
  Z_{U(1)}(q)=q^{\epsilon_0}\prod_{n=1}^\infty\frac{1}{1-q^n}\ .
\end{equation}
Here $\epsilon_0$ is the index Casimir energy contribution from the $U(1)$ degrees.
This zero point energy is given by $\epsilon_0=\frac{1}{2}{\rm tr}[(-1)^F(\epsilon-R_1)]$.
This contribution can be calculated from the letter index $z(q)=\frac{q}{1-q}$, as we
review in appendix B (which is explained in detail in \cite{Kim:2009wb}).
The result is
\begin{equation}
  \epsilon_0=\frac{1}{2}\lim_{q\rightarrow 1^-}q\frac{d}{dq}z(q)=
  \frac{1}{2\beta^2}-\frac{1}{24}\ ,
\end{equation}
where $q=e^{-\beta}$. After renormalization of the first divergent factor, one
obtains $\epsilon_0=-\frac{1}{24}$. This is basically the same as the zeta function
regularization, as the value of $\epsilon-R_1$ from the degrees in the letter index $z$
is $1,2,3,\cdots$. The zeta function regularization yields
$\frac{1}{2}\sum_{n=1}^\infty n=-\frac{1}{24}$. Inserting this in (\ref{free-index}),
the index becomes the inverse of the Dedekind eta function $\eta(\tau)$, where
$\tau$ is given by $q=e^{2\pi i\tau}$.

Thus, for our 5d approach to have any chance to capture the `free' $U(1)$ partition
function, or the partition function for the decoupled degrees coming from overall $U(1)$,
we should be able to find from a 5d calculation a multiplicative factor
$\frac{1}{\eta(\tau)}$. Using the modular property of $\eta(\tau)$,
one obtains the following expansion for small $\beta$:
\begin{equation}\label{overall-weak}
  Z_{U(1)}=\left(\frac{\beta}{2\pi}\right)^{\frac{1}{2}}
  e^{\frac{\pi^2}{6\beta}}\prod_{k=1}^\infty\frac{1}{1-e^{-\frac{4\pi^2k}{\beta}}}\ .
\end{equation}
This takes the form of a non-perturbative expansion in $\beta$.

Motivated by the above findings,
let us first consider what kind of corrections can appear to our 5d partition function.
From the saddle point equations (\ref{saddle}), Yang-Mills instanton configurations are
allowed on the $\mathbb{CP}^2$ base of $S^5$ in Hopf fibration. In our normalization for
$g_{YM}$, the classical action for $k$ instantons on the $\mathbb{CP}^2$ base is given
by\footnote{If we call the instantons of our saddle points to be self-dual, the Kahler
2-form $J_{\mu\nu}$ of $\mathbb{CP}^2$ is anti-self-dual. So the embedding of $J$ into
Abelian subgroup as $F_{\mu\nu}\sim J_{\mu\nu}$ \cite{Witten:2003ya} is excluded in our problem.}
\begin{equation}
  \frac{1}{4g_{YM}^2}\int_{\mathbb{CP}^2}{\rm tr}(F_{\mu\nu}F^{\mu\nu})
  =\frac{4\pi^2k}{g_{YM}^2}\ .
\end{equation}
This naturally yields the relation $\frac{4\pi^2}{g_{YM}^2}=\frac{1}{r_1}=\frac{2\pi}{r\beta}$
with $\beta\equiv\frac{2\pi r_1}{r_5}$. We introduced this in the introduction and also
used it in section 3.1. Despite the absence
of the physical D0-brane particle picture, we are suggesting that Euclidean D0-brane
loops which wraps a (possibly contractible) cycle, which we formally regard as time,
would provide the Kaluza-Klein `momentum' (in the sense of Fourier wavenumber) along
the extra circle. More precisely, the Euclidean D0-brane (or instanton)
action on $S^5$ is
\begin{equation}
  S_0=\frac{4\pi^2 k}{g_{YM}^2}\cdot 2\pi r=\frac{4\pi^2k}{\beta}\ .
\end{equation}
$2\pi r$ comes from the integration of the Lagrangian over the Hopf fiber direction $y$.
So the non-perturbative correction should take the form of
\begin{equation}\label{instanton-general}
  Z=\sum_{k=0}^\infty Z_ke^{-\frac{4\pi^2 k}{\beta}}\ ,
\end{equation}
which fits completely well with (\ref{overall-weak}), apart from the prefactor
$e^{\frac{\pi^2}{6\beta}}$ and $\left(\frac{\beta}{2\pi}\right)^{\frac{1}{2}}$.

To explain the last two factors, let us first turn to $e^{\frac{\pi^2}{6\beta}}$.
The presence of this factor can be understood by noticing that there could be a constant
shift to the supersymmetric actions on $S^5$ without modifying any symmetry. For instance,
\cite{Vafa:1994tf} emphasized in the context of topologically twisted 4d $\mathcal{N}=4$
SYM that there could be couplings of $g_{YM}$ ($\sim\beta$ in our case) to the background
curvature. In our case, on $S^5$, we may have constant couplings like
\begin{equation}
  \frac{\alpha}{g_{YM}^2}\int_{S^5}d^5x\sqrt{g}R^2\ ,
\end{equation}
where $R$ is the Riemann scalar curvature of $S^5$ and $\alpha$ is a
dimensionless constant. With a suitable coefficient $\alpha$, this term provides
the factor $e^{\frac{\pi^2}{6\beta}}$. As we have our freedom (or ambiguity) in 5d to
choose our theory on $S^5$, without spoiling any 5d symmetry, we implicitly assume
a certain constant shift of the action of the above form, so that the desired factor
comes out. As we are assuming the completeness of 5d SYM description, at least in
the BPS sector, such curvature couplings are restricted to $\mathcal{O}(\beta^{-3})$,
$\mathcal{O}(\beta^{-2})$, $\mathcal{O}(\beta^{-1})$ in general. So this is fixing a mild
ambiguity to get much more information on the 6d physics.

Now we turn to $\left(\frac{\beta}{2\pi}\right)^{\frac{1}{2}}$. We first note that
the perturbative partition function (\ref{perturbative}) at $N=1$ is just
$\left(\frac{\beta}{2\pi}\right)^{\frac{1}{2}}$. We take this factor from the perturbative
part and combine it with the instanton contribution of the form
(\ref{instanton-general}), to provide a desired factor in (\ref{overall-weak}).
Multiplying this factor from the perturbative part, now the non-perturbative series
(\ref{instanton-general}) takes the form of an index, supposing that the coefficients are
chosen to make (\ref{overall-weak}). So we find that, even for the $U(1)$ theory, the
structure of perturbative/non-perturbative contributions to the $S^5$ partition function
confronts and passes quite a nontrivial consistency test for it be an index.

Let us emphasize at this point that Abelian instantons, which we expect to account for
(\ref{overall-weak}), are not completely well defined purely within field theories,
as they often come with zero sizes which should be regarded as singular instantons.
In fact, non-Abelian instantons (at least in flat space) also have singularities in their
moduli spaces which correspond to small instantons. However, small instantons are often
important to understand various issues in string theory \cite{Witten:1995gx}. Often, giving
non-commutativity to the field theory makes the instanton moduli space smooth, and also makes Abelian instantons to be regular field theory solitons \cite{Nekrasov:1998ss}. This may
correspond to a (perhaps mild) UV completion of the 5d quantum field theory.

With these motivations, we now turn to the non-Abelian instanton corrections.
We only discuss the case with $U(N)$ gauge group. We claim that the full $U(N)$ non-perturbative partition function takes
the form
\begin{equation}
  Z(\beta)=Z_{\rm pert}(\beta)Z_{\rm inst}(\beta)\ ,
\end{equation}
where $Z_{\rm pert}$ is given in the previous subsection, and
\begin{equation}\label{U(N)-instanton}
  Z_{\rm inst}=\left[Z_{\rm inst}^{U(1)}\right]^N=
  e^{\frac{N\pi^2}{6\beta}}\prod_{k=1}^\infty\frac{1}{\left(1-e^{-\frac{4\pi^2k}{\beta}}
  \right)^N}=\eta(\tau)^{-N}
\end{equation}
with $\tau\equiv\frac{2\pi i}{\beta}$ (namely, $e^{2\pi i\tau}\equiv e^{-\frac{4\pi^2}{\beta}}$).
(\ref{U(N)-instanton}) takes
the general form of (\ref{instanton-general}), again with a suitable coupling to
the background curvature for the $e^{\frac{N\pi^2}{6\beta}}$ factor.
The proof of (\ref{U(N)-instanton}) will be presented in \cite{Kim:2012qf}, with
generalized to the squashed $S^5$. In this paper, we shall present several nontrivial
evidences and implications of this result.

Before studying the physics of (\ref{U(N)-instanton}), let us note
that the instanton partition functions are usually very simple in theories with $16$ SUSY.
In many important examples, the partition functions are
either $1$ or just functions of the coupling $g_{YM}^2$. On $\mathbb{R}^4$ and
$\mathbb{R}^4\times S^1$, \cite{Nekrasov:2002qd,Nekrasov:2003rj} calculates the instanton
partition function in the so-called Omega deformation $\epsilon_1$, $\epsilon_2$, which
roughly speaking compactifies the non-compact $\mathbb{R}^4$ part of the instanton moduli. When we consider the instanton partition function of maximal SYM, the following
simplifications appear. Although the instanton partition function depend on the VEV of a
scalar (similar to our saddle point value for $\phi$) in generic gauge theories with $8$
SUSY,  this dependence disappears at some special values of $\epsilon_1,\epsilon_2$
with $16$ SUSY. To explain some important cases, we first note that when
$\epsilon_1\!=\!\epsilon_2$, the instanton partition function just becomes $1$. This was a
crucial element in showing that the $S^4$ partition function for the $\mathcal{N}=4$ SYM
becomes a Gaussian matrix model with
$Z_{\rm pert}=Z_{\rm inst}=1$ \cite{Pestun:2007rz}. On the other hand, with
anti-self-dual Omega background with $\epsilon_1=-\epsilon_2\equiv\hbar$, the instanton
partition function for the $\mathcal{N}=4$ theory becomes independent of the remaining $\hbar$, and depends on $g_{YM}^2$ only. The partition function on the anti-self-dual
Omega background becomes \cite{Nekrasov:2003rj}
\begin{equation}
  Z_{\rm inst}=\frac{1}{\eta(\tau)^N}\ \ \ \ \ \ \left(
  \tau=\frac{\theta}{2\pi}+\frac{4\pi^2 i}{g_{YM}^2}\right)\ ,
\end{equation}
apart from the possible overall shift for the instanton number in the topologically
trivial sector, like those we discussed above. In particular, the result is the same
both for instantons on $\mathbb{R}^4$ or $\mathbb{R}^4\times S^1$. The instanton correction
(\ref{U(N)-instanton}) we propose on $S^5$ is basically the same as the result on
$\mathbb{R}^4$ or $\mathbb{R}^4\times S^1$, in anti-self-dual Omega background. The
relevance of these simpler cases to (\ref{U(N)-instanton}) is explained in \cite{Kim:2012qf}.
In this paper, we collect some evidences in favor of (\ref{U(N)-instanton}) and
discuss its physical implications.

Firstly, this yields the desired index (\ref{overall-weak}) for $N=1$.

Secondly, the nonperturbative result (\ref{U(N)-instanton}) can be dualized
for $\beta\gg 1$ to
\begin{equation}\label{instanton-dual}
  Z_{\rm inst}=\left(\frac{2\pi}{\beta}\right)^{N/2}\eta(i\beta/2\pi)^{-N}\ ,
\end{equation}
using the S-dual modular property of the eta function. The factor $\left(\frac{\beta}{2\pi}\right)^{-N/2}$ in (\ref{instanton-dual}) combines with
a factor $\left(\frac{\beta}{2\pi}\right)^{N/2}$ in the perturbative partition function
(\ref{perturbative}) which prevents an index interpretation of (\ref{perturbative}).
Moving it to the non-perturbative part and combining this with (\ref{U(N)-instanton}),
one finds that both perturbative and non-perturbative parts take the form of an index,
since
\begin{equation}
  \left(\frac{\beta}{2\pi}\right)^{N/2}Z_{\rm inst}=\frac{1}{\eta(i\beta/2\pi)^N}
  =q^{\frac{N}{24}}\prod_{n=1}^N\frac{1}{(1-q^n)^N}\ ,
\end{equation}
where we defined $q\equiv e^{-\beta}$.
So the structure (\ref{instanton-general}) of instanton expansion conspires well
with the provided prefactor in the perturbative part, to make the whole expression
$Z_{\rm pert}Z_{\rm inst}$ an index. It is somewhat curious to find that the
perturbative and non-perturbative parts have to combine for the 5d SYM to tell us the 6d
physics consistently.\footnote{This sounds a bit similar to the failure of perturbative
finiteness of maximal SYM \cite{Bern:2012di}. The only chance for this theory to be UV
complete is then by combining the full perturbative/non-perturbative effects at the
cutoff scale where the distinction between the two becomes meaningless \cite{Lambert:2012qy}. Although we do not see any serious divergence in our SUSY path integral,
the consistency of 6d physics still requires us to combine to two.}

Most importantly, we shall now show that the non-perturbative completion (\ref{U(N)-instanton}) perfectly agrees with the large $N$ index that we know from
the gravity dual on $AdS_7\times S^4$. Before combining the instanton correction (\ref{U(N)-instanton}), the perturbative
part (\ref{perturbative}) shows a very strange large $N$ behavior. Let us consider the
part which gives the degeneracy information:
\begin{equation}\label{pert-expand}
  \prod_{n=1}^{N-1}(1-q^n)^{N-n}=1-(N-1)q^1+\frac{N^2-5N+6}{2}q^2
  -\frac{N^3-12N^2+35N-36}{6}q^3+\cdots\ .
\end{equation}
The low energy degeneracy at large $N$ is so large that this part alone will not have
a sensible large $N$ limit: especially it cannot have a large $N$ gravity dual on $AdS$,
which exhibits a low energy spectrum which is completely independent of $N$.
Combining $Z_{\rm inst}$ with the perturbative contribution, one obtains
\begin{equation}\label{full-index}
  Z=Z_{\rm pert}Z_{\rm inst}=e^{\frac{N(N^2-1)\beta}{6}}\prod_{n=1}^{N-1}(1-e^{-n\beta})^{N-n}
  \cdot e^{\frac{N\beta}{24}}\prod_{n=1}^\infty\frac{1}{(1-e^{-n\beta})^N}\ .
\end{equation}
The large $N$ index, apart from the zero point
energy part, is given by the MacMahon function,
\begin{equation}\label{large-N-QFT}
  Z_{N\rightarrow\infty}=\prod_{n=1}^\infty\frac{1}{(1-q^n)^n}\ .
\end{equation}
Again, we used $q\equiv e^{-\beta}$.
We see that the contribution of $\mathcal{O}(N)$ fermionic `letters' at low energy
in (\ref{pert-expand}) cancels with the $\mathcal{O}(N)$ bosonic letter contributions,
leaving $\mathcal{O}(1)$ low energy degeneracy.

\begin{table}[t!]
$$
\begin{array}{c|ccc|c}
  \hline &\epsilon&SO(6)&SO(5)&{\rm boson/fermion}\\
  \hline p\geq 1&2p&(0,0,0)&(p,0)&{\rm b}\\
  p\geq 1&2p+\frac{1}{2}&(\frac{1}{2},\frac{1}{2},\frac{1}{2})
  &(p-\frac{1}{2},\frac{1}{2})&{\rm f}\\
  p\geq 2&2p+1&(1,0,0)&(p-1,1)&{\rm b}\\
  p\geq 3&2p+\frac{3}{2}&(\frac{1}{2},\frac{1}{2},-\frac{1}{2})
  &(p-\frac{3}{2},\frac{3}{2})&{\rm f}\\
  \hline \cdot&\frac{7}{2}&(\frac{1}{2},\frac{1}{2},-\frac{1}{2})
  &(\frac{1}{2},\frac{1}{2})&{\rm b\ (fermionic\ constraint)}\\
  \hline
\end{array}
$$
\caption{BPS Kaluza-Klein fields of $AdS_7\times S^4$ supergravity}\label{sugra}
\end{table}
Now we study the same index in the large $N$ limit from the $AdS_7\times S^4$
supergravity, giving the weight $q^{\epsilon-R_1}$ to the low energy gravity states.
Again choosing a particular $Q$, $S$ and viewing our index as the unrefined version
of the superconformal index
associated with $Q,S$, it suffices for us to consider the contribution from gravity
states preserving these SUSY. The Kaluza-Klein field contents are given in
\cite{Bhattacharya:2008zy}, and we only list the BPS fields in Table \ref{sugra}.
Collecting all the contributions, one obtains the single particle gravity index
\begin{eqnarray}
  I_{\rm sp}(q)&=&\frac{1}{(1-q)^3}\left[\sum_{p=1}^\infty\sum_{n=0}^pq^{2p-n}
  -3\sum_{p=1}^\infty\sum_{n=1}^pq^{2p+1-n}+3\sum_{p=2}^\infty\sum_{n=1}^{p-1}q^{2p+1-n}
  -\sum_{p=3}^\infty\sum_{n=1}^{p-2} q^{2p+1-n}+q^3\right]\nonumber\\
  &=&\frac{q}{(1-q)^2}=q+2q^2+3q^3+4q^4+\cdots\ .
\end{eqnarray}
The multiparticle exponent of $I_{\rm sp}$ yields the MacMahon function
\begin{equation}
  I_{\rm mp}(q)=\exp\left[\sum_{n=1}^\infty\frac{1}{n}I_{\rm sp}(q^n)\right]
  =\prod_{n=1}^\infty\frac{1}{(1-q^n)^n}
\end{equation}
as the multiparticle gravity index on $AdS_7\times S^4$, precisely agreeing
with the result (\ref{large-N-QFT}) from 5d gauge theory calculation.

It is curious to find that the non-perturbative correction (\ref{U(N)-instanton})
yields an $\mathcal{O}(N)$ correction to the `index Casimir energy' obtained from
the perturbative part. One obtains
\begin{equation}
  \epsilon_0=(\epsilon_0)_{\rm pert}+(\epsilon_0)_{\rm inst}=
  -\frac{N(N^2-1)}{6}-\frac{N}{24}\ .
\end{equation}
It would be curious to see if this can be understood as a combination
of various anomaly coefficients of the 6d $(2,0)$ theory \cite{Harvey:1998bx},
similar to what we observe for the 4d Casimir energy in appendix B.

Finally, MacMahon function that we obtained at large $N$ is well-known as
the generating function for the 3 dimensional Young diagrams. Curiously, our finite
$N$ index (\ref{full-index}) is the generating function for the 3d Young diagrams
with their heights being no longer than $N$. This index also coincides with
the vacuum character of the $W_N$ algebra, apart from a factor of eta function
\cite{Gaberdiel:2010ar}. It should be interesting to seek for the physical meanings of
these apparently surprising relations, if any.\footnote{We thank Amihay Hanany for
discussions which led us to the observation on the restricted 3d Young
diagrams. Also, we thank Rajesh Gopakumar for explaining the coincidence with
the $W_N$ vacuum character.}

\subsection{Generalizations}

One can easily modify the localization calculus for the maximal SYM on $S^5$
to include two chemical potentials conjugate to $\epsilon-R_1$ and $\epsilon-R_2$.
One has to calculate the $S^5$ partition function for the theory with $8$ SUSY,
with two parameters $\beta\sim g_{YM}^2$ and $\Delta$.

By following the calculation similar to the case with $\Delta=1$ in section 3.1
and appendix A, one obtains similar cancelations between non-BPS modes and finds
the following integrand of the Hermitian matrix integral.
Firstly, the classical contribution and the determinant from the vector multiplet
part does not change compared to the analysis in the previous section. The
hypermultiplet contribution changes as
\begin{eqnarray}\label{hyper-det}
  &&\prod_{\alpha\in{\rm root}}\prod_{k=0}^\infty\left(k+1+\Delta+ir\alpha(\phi)
  \right)^{-\frac{(k+1)(k+2)}{2}}
  \left(k+2-\Delta+ir\alpha(\phi)\right)^{-\frac{(k+1)(k+2)}{2}}\nonumber\\
  &&=\prod_{\alpha\in{\rm root}}
  \prod_{k=1}^\infty\left(k-1+\Delta+ir\alpha(\phi)\right)^{-\frac{k^2-k}{2}}
  \left(k+1-\Delta+ir\alpha(\phi)\right)^{-\frac{k^2+k}{2}}\ .
\end{eqnarray}
Notice that our previous partition function at $\Delta=1$ is same as that with
$\Delta=0$, as the two points just exchange the roles of $R_1$ and $R_2$.\footnote{This
essentially gives the determinant for a hypermultiplet in a real representation, if one replaces $\alpha(\phi)$ in (\ref{hyper-det}) by $\mu(\phi)$, where $\mu$ runs over
the weights in the representation. For a complex representation, one has to multiply
a similar factor with $\mu(\phi)$ replaced by $-\mu(\phi)$, and then take square root to
get the determinant \cite{Kim:2012qf}.} The full
integrand, apart from the Gaussian factor and the Vandermonde measure, is given by
\begin{equation}\label{general-determinant}
  \prod_{\alpha\in{\rm root}}
  \prod_{k=1}^\infty\frac{(k+ir\alpha(\phi))^{k^2+2}}
  {\left(k-1+\Delta+ir\alpha(\phi)\right)^{\frac{k^2-k}{2}}
  \left(k+1-\Delta+ir\alpha(\phi)\right)^{\frac{k^2+k}{2}}}\ .
\end{equation}
The exact integration with Gaussian measure and the Vandermonde determinant
does not seem to be as simple as our previous example with $16$ SUSY.

The above infinite product requires regularization. Various factors in
(\ref{general-determinant}) are all regularized in the literatures using zeta
function regularization. One obtains (we use the fact that adjoint representation
is real to obtain the second line)
\begin{eqnarray}
  &&\prod_{\alpha\in{\rm root}}
  \prod_{k=1}^\infty(k+ir\alpha(\phi))^2\cdot\frac{(k+ir\alpha(\phi))^{k^2}}
  {\left(k-1+\Delta+i\alpha(\phi)\right)^{\frac{k^2}{2}}
  \left(k+1-\Delta+i\alpha(\phi)\right)^{\frac{k^2}{2}}}\cdot
  \left(\frac{k-1+\Delta+ir\alpha(\phi)}{k+1-\Delta+ir\alpha(\phi)}\right)^{\frac{k}{2}}\nonumber\\
  &&\longrightarrow\prod_{\alpha\in{\rm root}}\frac{2\pi\sinh(\pi r\alpha(\phi))}{\pi r\alpha(\phi)}
  \cdot e^{\frac{1}{2}f(ir\alpha(\phi))-\frac{1}{2}f(1-\Delta+ir\alpha(\phi))}\cdot e^{-\frac{1}{2}\ell(1-\Delta+ir\alpha(\phi))}\ ,
\end{eqnarray}
where the function $f(x)$ (even in $x\rightarrow -x$) is given by \cite{Kallen:2012cs}
\begin{equation}
  f(x)=\frac{i\pi x^3}{3}+x^2(1-e^{-2\pi ix})+\frac{ix{\rm Li}_2(e^{-2\pi i x})}{\pi}
  +\frac{{\rm Li}_3(e^{-2\pi ix})}{2\pi^2}-\frac{\zeta(3)}{2\pi^2}\ ,
\end{equation}
and the (odd) function $\ell(x)$ is given by \cite{Jafferis:2010un}
\begin{equation}
  \ell(x)=\frac{i\pi x^2}{2}-x\log(1-e^{2\pi ix})
  +\frac{i{\rm Li}_2(e^{2\pi ix})}{2\pi}-\frac{i\pi}{12}\ .
\end{equation}
The matrix integral is given by
\begin{equation}
  \frac{1}{|W|}\int d\lambda e^{-\frac{2\pi^2{\rm tr}(\lambda^2)}{\beta}}
  \prod_{\alpha\in{\rm root}}2\sinh(\pi\alpha(\lambda))
  e^{\frac{1}{2}f(i\alpha(\lambda))-\frac{1}{2}f(1-\Delta+i\alpha(\lambda))
  -\frac{1}{2}\ell(1-\Delta+i\alpha(\lambda))}\ ,
\end{equation}
where $\lambda\equiv r\phi$.

We first note that the limit $\beta\rightarrow\infty$, $\Delta\rightarrow 2$ with
fixed $\beta(2-\Delta)>0$, towards the half-BPS partition function, is quite singular
and may drastically change the nature of the matrix integral. Firstly, the strong coupling
limit $\beta\rightarrow\infty$ takes the Gaussian measure to $1$. Secondly, the second
term in the denominator, at $k=1$ yields a factor $1-\Delta+ir\alpha(\phi)$ which at $\Delta=2$
completely cancels the zeros in the $\sinh$ measure. So in this limit, there are no short
distance repulsions between different eigenvalues.
We still have a parameter $\beta_H=\beta(2-\Delta)$ which gives the fugacity $q=e^{-\beta_H}$
of the half-BPS partition function. One can thus consider calculating the matrix integral
in a series expansion of $\beta_H$, and compare with the expected half-BPS partition function.
As $(q_1q_2)^{\epsilon_0}$ becomes infinity with negative $\epsilon_0$, we expect to have
a divergent prefactor multiplying the conventional half-BPS partition function.

For simplicity, let us consider the $U(N)$ half-BPS partition function.
The $U(N)$ partition function for half-BPS states is given by \cite{Bhattacharyya:2007sa}
\begin{equation}\label{half-BPS}
  Z=\prod_{n=1}^N\frac{1}{1-q^n}\ ,
\end{equation}
up to a divergent multiplicative factor, with $q=e^{-\beta_H}$ as defined in the
previous paragraph. It will be interesting to see whether our result, supplemented by the
instanton correction of \cite{Kim:2012qf}, reproduces (\ref{half-BPS}).

\section{Discussions}

In this paper, we explored the possibility that partition functions of SYM on $S^5$
could capture the indices of the 6d $(2,0)$ theory on $S^5\times S^1$. The 5d field theories
are carefully chosen, by first studying the Scherk-Schwarz reductions of Abelian $(2,0)$
theories on $S^5\times S^1$ on the circle, and then trying to generalize to non-Abelian theories on $S^5$. We showed that the partition function for the maximal SYM on $S^5$
captures the physics of the 6d $(2,0)$ theory in a surprisingly accurate and detailed manner.

Firstly, the partition function takes the form of an index, which from a naive 5d perspective
has no reason to be true. Generalizing the idea to other 5d theories on $S^5$, the requirement that the partition function take the form of an index could severely restrict the class of
theories having a 6d UV fixed point. For instance, it should be desirable to further study
the index for the $(2,0)$ theory with more complicated chemical potentials, from 5d gauge
theories with as little as $2$ SUSY (those preserved by the most refined superconformal
index). Also, studying our partition function for other gauge groups will also be interesting. One can also study a 5d reduction of the 6d $(1,0)$ superconformal theories.

We find that our index captures the $N^3$ some of degrees of freedom by what we called
the `index Casimir energy,' which is a Casimir energy like quantity appearing in
the index. It should be interesting to see if this is an observable which is worth further
studies. Also, possible relations to other suggested measures of degrees of freedom could
be interesting. Derivation of our index Casimir energy from the gravity dual should
also be very important.

We showed that the index calculated
from the 5d maximal SYM with $U(N)$ gauge group completely agrees with the supergravity
index on $AdS_7\times S^4$ in the large $N$ limit. We find this as quite a nontrivial
signal that our approach is on the right way. Similar successful matching
of instanton partition function on $\mathbb{R}^4\times S^1$ and the DLCQ supergravity
index on $AdS_7\times S^4$ was found in \cite{Kim:2011mv}.

The perturbative partition function that we find for maximal SYM on $S^5$ turns out
to be identical to the pure Chern-Simons partition function on $S^3$. Possible physical
connections between the two observables are not clear to us at the moment.
However, inspired by the fact that the Jones polynomial and other topological invariants
were studied by Wilson loop observables in Chern-Simons theories \cite{Witten:1988hf},
one may ask if the Wilson loops in 5d gauge theories can play interesting roles as well.
Earlier works on Wilson loops in 5d SYM include \cite{Young:2011aa}.

\vskip 0.5cm

\hspace*{-0.8cm} {\bf\large Acknowledgements}

\vskip 0.2cm

\hspace*{-0.75cm} We are grateful to Dongmin Gang, Eunkyung Koh, Kimyeong Lee, Sangmin Lee,
Costis Papageorgakis, Jaemo Park, Jeong-Hyuck Park and Shuichi Yokoyama for discussions, and
especially to Jungmin Kim for collaboration and comments. S.K. thanks Sunil Mukhi and
Alessandro Tomasiello for discussions on related subjects, and Marcos Marino for discussions
on Casimir energies a while ago. We also thank many valuable comments on the first
version of this paper, especially those from Rajesh Gopakumar, Amihay Hanany,
Kazuo Hosomichi, Daniel Jafferis, Igor Klebanov,  Costis Papageorgakis, Leonardo Rastelli,
Yuji Tachikawa, Martin Wolf and Maxim Zabzine. This work is supported by the BK21
program of the Ministry of Education, Science and Technology (SK), the National Research
Foundation of Korea (NRF) Grants No. 2010-0007512 (HK, SK), 2012R1A1A2042474 (SK),
2012R1A2A2A02046739 (SK) and 2005-0049409 through the Center for Quantum Spacetime
(CQUeST) of Sogang University (SK).

\appendix

\section{Spinors, spherical harmonics and determinants}

In this paper, we mostly view $S^5$ as a Hopf fibration over $\mathbb{CP}^2$.
The metric of a round sphere with unit radius is given by
\begin{eqnarray}
  ds^2(S^5)&=&ds^2(\mathbb{CP}^2)+\left(dy+\frac{1}{2}\sin^2\rho\sigma^3\right)^2\\
  ds^2(\mathbb{CP}^2)&=&d\rho^2+\frac{1}{4}\sin^2\rho(\sigma_1^2+\sigma_2^2)+
  \frac{1}{4}\sin^2\rho\cos^2\rho\ \sigma_3^2\equiv e^ae^a\nonumber
\end{eqnarray}
with the vierbein
\begin{equation}
  e^1=d\rho\ ,\ \ e^2=\frac{1}{2}\sin\rho\cos\rho\ \sigma_3\ ,\ \
  e^3=\frac{1}{2}\sin\rho\ \sigma_1\ ,\ \ e^4=\frac{1}{2}\sin\rho\ \sigma_2
\end{equation}
on $\mathbb{CP}^2$ and the 1-forms
\be
    \sigma^1 &=& \sin\psi d\theta -\cos\psi\sin\theta d\phi \ , \nn \\
    \sigma^2 &=& \cos\psi d\theta +\sin\psi\sin\theta d\phi \ , \nn \\
    \sigma^3 &=& d\psi + \cos\theta d\phi\ .
\ee

We first consider the spherical harmonics on $S^5$ that we use.
Let us start with the scalar spherical harmonics. We denote the scalar harmonics
by $Y^{k}$ which are defined as eigenfunctions of the Laplacian on $S^5$
satisfying the eigenvalue equation
\be
    -\nabla^\mu \nabla_\mu Y^k = k(k+4) Y^k \ , \quad k\ge 0 \ ,
\ee
with degeneracy $\frac{1}{12}(k+1)(k+2)^2(k+3)$.
They take a representation of the $SO(6)$ isometry group on $S^5$.
Then one can further decompose the harmonics by the eigenvalues of
one of $SO(6)$ Cartan generators such as
\be
   L_v Y^k = v\cdot \nabla Y^k= im Y^k \ , \quad (m=-k,-k+2,\cdots ,k-2,k)\ .
\ee
Modes with given $k,m$ have degeneracy $\frac{1}{8}(k+2)\left((k+2)^2-m^2\right)$.
$L_v$ is the Lie derivative along the Hopf fiber of $S^5$.

The spinor harmonics can be constructed using the scalar spherical harmonics $Y^k$
with simple Killing spinors $\eta_\pm$ on $S^5$ (which we call $\epsilon^\pm$ in
section 3) satisfying
\be
    \nabla_\mu\eta_\pm = \pm \frac{i}{2r}\gamma_\mu\eta_\pm \ , \quad \gamma^{12}\eta_\pm = -\gamma^{34}\eta_\pm = \pm i \eta_\pm \ .
\ee
These spinors are normalized as $\bar\eta_+ \eta_+ = \bar\eta_-\eta_-=1$
and their bilinear produces the vector along Hopf fiber direction
$\bar\eta_+\gamma^\mu\eta_+ = v^{\mu}$.
One can choose the following basis for the spinor harmonics (inspired by \cite{Pope:1980ub})
\be
    \begin{array}{lc}\Psi_1 = Y^k\eta_+ &  , \ -k \le m \le k \\
     \Psi_2 = \gamma^\mu \hat{\nabla}_\mu Y^k \eta_+ & , \ -k\le m< k \\
     \Psi_3 = Y^k \eta_- & , \ -k \le m \le k \\
     \Psi_4 = \gamma^\mu \hat{\nabla}_\mu Y^k\eta_- & , \ -k < m \le k\end{array}\ .
\ee
where $\hat{\nabla}_\mu = \nabla_\mu - v_\mu v\cdot\nabla$ is the projected derivative on to
$\mathbb{CP}^2$ base.
Note that, as $\eta_\pm$ has $\gamma^5$  eigenvalue $+1$ and the operator $\gamma^\mu \hat{\nabla}_\mu$ anticommutes with $\gamma^5$,
the basis $\Psi_2$ and $\Psi_4$ have $\gamma^5$ eigenvalues $-1$.
Therefore the above basis $\Psi$ span the complete basis of the spinor harmonics.
Then it is straightforward to find the linear combinations of the spinor basis $\Psi$
to form the eigenfunctions of the Dirac equation on $S^5$.
The Dirac operator acts on $\Psi$'s as
\be
    &&\left\{\begin{array}{l}\gamma^\mu\nabla_\mu \Psi_1 = i(\frac{5}{2}+m) \Psi_1 + \Psi_2 \\
    \gamma^\mu\nabla_\mu \Psi_2 = -(k-m)(k+m+4)\Psi_1 -i(\frac{3}{2}+m)\Psi_2\end{array}\right. \nn \\
    &&\left\{\begin{array}{l}\gamma^\mu\nabla_\mu \Psi_3 = -i(\frac{5}{2}-m) \Psi_3 + \Psi_4 \\
    \gamma^\mu\nabla_\mu \Psi_4 = -(k+m)(k-m+4)\Psi_3 +i(\frac{3}{2}-m)\Psi_4\ .\end{array}\right.
\ee
The eigenvalues of the Dirac operator are then given by
\be
    i\gamma^\mu\nabla_\mu \Psi \rightarrow \left\{ \begin{array}{cl} +k +\frac{5}{2} & \\
    -k-\frac{5}{2} & \\ +k+\frac{3}{2} & , \ m \neq k \\ -k-\frac{3}{2} & , \ m \neq -k\ .\end{array}  \right.
\ee
Thus we can rearrange the spinor harmonics into the two set of eigenstates with the eigenvalues $\pm(k + \frac{5}{2})$ and the degeneracy $\frac{1}{6}(k+1)(k+2)(k+3)(k+4)$
where $k \ge 0$.

For the vector spherical harmonics, we first study the eigenfunctions of the Maxwell
operator on $S^5$:
\be
    (-\nabla^2 \delta^\nu_\mu +\nabla_\mu\nabla^\nu +4\delta^\nu_\mu)A_\nu
    = \mathcal{O}_\mu^\nu \mathcal{A}_\nu\ .
\ee
The spinor basis $\Psi$ can be used to construction the basis for the vector harmonics.
Let us introduce the divergenceless vector basis
\be
    \begin{array}{ll}
        \mathcal{A}^1_\mu = \eta^\dagger_+ \gamma_\mu \Psi_1 +\frac{im}{k(k+4)}
        \nabla_\mu Y^k= v_\mu Y^k +\frac{im}{k(k+4)}\nabla_\mu Y^k&   \\
        \mathcal{A}^2_\mu = \eta^\dagger_+ \gamma_\mu \Psi_2 -\left(1-\frac{m(m+4)}{k(k+4)}\right)\nabla_\mu Y^k & \\
        \mathcal{A}^3_\mu = \eta^\dagger_+ \gamma_\mu \Psi_4 & \\
        \mathcal{A}^4_\mu = \eta^\dagger_-\gamma_\mu \Psi_2 &
    \end{array}
\ee
where they satisfy the Lorentz gauge condition $\nabla^\mu\mathcal{A}_\mu = 0$.
The factors $\nabla_\mu Y^k$ are added to $\mathcal{A}_\mu^1$ and $\mathcal{A}_\mu^2$
for this.
In this construction, one can also consider other basis such as $\eta^\dagger_+\gamma_\mu \Psi_3$ but they identically vanish due to the projection properties of $\eta_\pm$.
The vectors $\mathcal{A}^{1,2,3,4}$ together with the pure gauge $\nabla_\mu Y^k$
constitutes the 5 basis of the vector spherical harmonics.
The Maxwell operator $\mathcal{O}_\mu^\nu$ acts on $\mathcal{A}_\mu$ as
\be\label{Maxwell-eigen}
    &&\left\{\begin{array}{l}\mathcal{O}_\mu^\nu \mathcal{A}^1_\nu = \left(k(k+4)+2m+8\right)\mathcal{A}^1_\mu-2i\mathcal{A}^2_\mu \\
    \mathcal{O}_\mu^\nu \mathcal{A}^2_\nu =2i\left(k(k+4)-m(m+4)\right)\mathcal{A}^1_\mu+\left(k(k+4)-2m\right)\mathcal{A}^2_\mu \end{array} \right. \nn \\
    &&\hspace{.5cm} \mathcal{O}_\mu^\nu \mathcal{A}^3_\nu = (k+1)(k+3)\mathcal{A}^3_\mu \nn \\
     &&\hspace{.5cm} \mathcal{O}_\mu^\nu \mathcal{A}^4_\nu = (k+1)(k+3)\mathcal{A}^4_\mu
\ee
while its action on the pure gauge $\nabla_\mu Y^k$ is trivial.
The eigenvalues are given by
\be
    \mathcal{O}_\mu^\nu\mathcal{A}_\nu \rightarrow \left\{\begin{array}{cl}
    (k+2)(k+4) & , \ k\ge 0 \\
    k(k+2) & , \ k\ge2 \ {\rm and} \ m\neq \pm k \\
    (k+1)(k+3) & , \ k \ge 1 \ {\rm and}\ m\neq k \ {\rm or} \ -k \end{array}\right.
\ee
They can also be rearranged to the vector harmonics having eigenvalue $(k+2)(k+4)$ with degeneracy $\frac{1}{3}(k+1)(k+3)^2(k+5)$ where $k\ge0$.

We now compute one-loop determinant of the quadratic action in the $\mathcal{Q}$-exact deformations.
We first focus on the vector multiplet.
The $\mathcal{Q}$-exact deformations is given in eqn (3.5).
The integrals over the fluctuations of the auxiliary scalars $D^I$ are trivial and the contribution from the scalar field $\phi$ is canceled with the pure gauge and the ghost contributions.
For the gauge part, we need to diagonalize the quadratic terms
\be
	&&\frac{1}{2}F_{\mu\nu}F^{\mu\nu} -\frac{1}{4} \epsilon^{\mu\nu\lambda\rho\sigma}v_\mu F_{\nu\lambda}F_{\rho\sigma} -[A_\mu,\phi_0]^2\nn \\
    &=& A^\mu\left(-\nabla^2 \delta_\mu^\nu + \nabla_\mu\nabla^\nu + 4\delta_\mu^\nu -2(J_{\mu\lambda}v\cdot\nabla + 2v_{[\mu}J_{\lambda]\rho}\nabla^\rho) g^{\lambda\nu}\right)A_\nu -[A_\mu,\phi_0]^2 \nn \\
    &\equiv& A^\mu (\mathcal{O}_\mu^\nu + \hat{\mathcal{O}}_\mu^\nu )A_\nu -[A_\mu,\phi_0]^2
\ee
The vector basis $\mathcal{A}$'s can be used to diagonalized these terms.
We obtain
\be
	&&\left\{\begin{array}{l}\hat{\mathcal{O}}_\mu^\nu \mathcal{A}^1_\nu = 2(m+4)\mathcal{A}^1_\mu -2i\mathcal{A}^2_\mu \\
	\hat{\mathcal{O}}_\mu^\nu\mathcal{A}^2_\nu = 2i\left(k(k+4)-m(m-4)\right)\mathcal{A}^1_\mu -2m \mathcal{A}^2_\mu
	\end{array}\right. \nn \\
	&& \hspace{.5cm}\hat{\mathcal{O}}_\mu^\nu \mathcal{A}^{3}_\nu = 2(3-m)\mathcal{A}^{3}_\mu \nn \\
	&& \hspace{.5cm}\hat{\mathcal{O}}_\mu^\nu \mathcal{A}^{4}_\nu = 2(3+m)\mathcal{A}^{4}_\mu
\ee
Plugging these results with \eqref{Maxwell-eigen}, the eigenvalues of the gauge part becomes
\be
	\left\{\begin{array}{cl} (k+4)^2 + \alpha(\phi_0)^2 & , \ k\ge0\\
	k^2 +\alpha(\phi_0)^2& , \ k\ge 2 \ {\rm and} \ m\neq \pm k \\
	k(k+4)-2m+9+\alpha(\phi_0)^2 & , \ k\ge 1 \ {\rm and} \ m\neq -k \\
	k(k+4)+2m+9+\alpha(\phi_0)^2 & , \ k\ge 1 \ {\rm and} \ m\neq k\end{array}\right.
\ee
This leads to the bosonic one-loop determinant from the vector multiplet
\be\label{1-loop-gauge-boson}
{\rm det}_{V,b}\!\!&\!\!=\!\!&\!\!\prod_{\alpha\in root}\left[\prod_{k=0}^\infty\big((k+4)^2+\alpha(\phi_0)^2\big)^{\frac{1}{24}(k+1)(k+2)^2(k+3)}\prod_{k=2}^\infty\big(k^2+\alpha(\phi_0)^2\big)^{\frac{1}{24}(k+1)(k+2)^2(k+3)-\frac{(k+1)(k+2)}{2}} \right. \nn \\
	 &&\hspace{1cm}\left.\times\prod_{k=1}^\infty\prod_{m=-k+1}^k\big(k(k+4)-2m+9+\alpha(\phi_0)^2\big)^{\frac{1}{8}(k+2)\left((k+2)^2-m^2\right)}\right]
\ee
We then turn to the fermionic contribution. The quadratic terms for the gaugino are
\be
-i\chi^\dagger\gamma^\mu\nabla_\mu\chi-i[\phi,\chi^\dagger]\chi+\chi^\dagger\sigma^3\chi-\frac{1}{2}\chi^\dagger \gamma_5\sigma^3\chi-\frac{i}{2}\chi^\dagger(\gamma^{12}-\gamma^{34})\chi\ .
\ee
We will consider only a complex spinor field $\chi^{A=1}$ in the doublet of $SU(2)_R$ as the other field $\chi^{A=2}$ is related to $\chi^1$ by
the reality condition.
Then the fermionic quadratic terms reduce to
\be
    \chi^\dagger\left[-i\gamma^\mu\nabla_\mu+i\alpha(\phi_0)+1-\frac{1}{2}(\gamma^5+i\gamma^{12}-i\gamma^{34})\right]\chi \equiv \chi^\dagger\mathcal{O}_{V,f}\chi\ .
\ee
Using the spinor harmonics $\Psi$, one can easily show that this quadratic operator acts on $\Psi$ as
\be
	&&\left\{\begin{array}{l} \mathcal{O}_{V,f}\Psi_1=\big(m+4+i\alpha(\phi_0)\big) \Psi_1 -i\Psi_2 \\
	\mathcal{O}_{V,f}\Psi_2= i(k-m)(k+m+4)\Psi_1+\big(\!-\!m+i\alpha(\phi_0)\big)\Psi_2
	\end{array}\right. \nn \\
	&&\left\{\begin{array}{l} \mathcal{O}_{V,f}\Psi_3=\big(m-3+i\alpha(\phi_0)\big)\Psi_3-i\Psi_4 \\
	\mathcal{O}_{V,f}\Psi_4= i(k+m)(k-m+4)\Psi_3+\big(\!-\!m+3+i\alpha(\phi_0)\big)\Psi_4
	\ .\end{array}
	\right.
\ee
By considering the proper degeneracy of the spinor basis $\Psi$,
we obtain the fermionic determinant of the vector multiplet
\be\label{1-loop-gauge-fermion}
{\rm det}_{V,f}\!\!&\!\!=\!\!&\!\!\prod_{\alpha\in root}
\left[\prod_{k=0}^\infty\big(k+4+i\alpha(\phi_0)\big)^{\frac{1}{12}(k+1)(k+2)^2(k+3)}
\big(k+i\alpha(\phi_0)\big)^{\frac{1}{12}(k+1)(k+2)^2(k+3)-\frac{(k+1)(k+2)}{2}} \right.\\ &&\hspace{1cm}\left.\times\prod_{k=0}^\infty\big(k+3+i\alpha(\phi_0)\big)^{\frac{(k+1)(k+2)}{2}}
\!\!\prod_{m=-k+1}^k\!\!\big(k(k+4)-2m+9+\alpha(\phi_0)^2\big)^{\frac{1}{8}(k+2)
\left((k+2)^2-m^2\right)}\right].\nonumber
\ee
There is a huge cancelation between the bosonic contribution \eqref{1-loop-gauge-boson}
and the fermionic contribution \eqref{1-loop-gauge-fermion}.
Collecting the remaining terms, the one-loop perturbative part of the path integral
for the vector multiplet becomes
\be
	\frac{{\rm det}_{V,f}}{{\rm det}_{V,b}}\!\!&\!\!=\!\!&\!\!\prod_{\alpha\in root}\prod_{k=0}^\infty\big(k+3+i\alpha(\phi_0)\big)^{\frac{(k+1)(k+2)}{2}}\prod_{k=1}^\infty\big(k+i\alpha(\phi_0)\big)^{\frac{(k+1)(k+2)}{2}} \nn \\
	&=&\prod_{\alpha\in root}\prod_{k=1}^\infty\big(k+i\alpha(\phi_0)\big)^{k^2+2}\ .
\ee

Let us move on to the hypermultiplet part.
The $\mathcal{Q}$-exact deformation generalized by a continuous parameter $\Delta$ is
\be
	 \hspace{-.7cm}&&\frac{1}{2}\delta\left((\delta\psi)^\dagger+\psi^\dagger(\delta\psi^\dagger)^\dagger\right) \nn \\
	\hspace{-.7cm}&&=|D_\mu q^A|^2-\frac{i}{r}v^\mu\bar{q}\sigma^3 D_\mu q+\frac{i}{r}(1-2\Delta)v^\mu \bar{q}D_\mu q
    +\frac{(\Delta-2)^2}{r^2}\bar{q}_1 q^1 +\frac{(\Delta+1)^2}{r^2}\bar{q}_2 q^2 +|[\phi_0,q^A]|^2 - \bar{F}_{A'}F^{A'} \nn \\
    \hspace{-.7cm}&&+i\psi^\dagger\gamma^\mu D_\mu \psi +\frac{1-2\Delta}{2r}v^\mu \psi^\dagger\gamma_\mu\psi -\frac{i}{4r}J^{\mu\nu}\psi^\dagger\gamma_{\mu\nu}\psi + i\psi^\dagger[\phi_0,\psi]\ .
\ee
The integrals of the auxiliary scalars $F^{A'}$ are trivial.
The matter scalar field $q^A$ gives the bosonic determinant
\be
	{\rm det}_{H,b}\!\!&\!\!=\!\!&\!\!\prod_{\alpha\in root}\prod_{k=0}^\infty\prod_{m=-k}^k\Big[
	\big(k(k+4)+2m\Delta+(\Delta-2)^2+\alpha(\phi_0)^2\big)^{\frac{1}{8}(k+2)((k+2)^2-m^2)}\\
    && \hspace{2cm} \times \big(k(k+4)+2m(\Delta-1)+(\Delta+1)^2+\alpha(\phi_0)^2\big)^{\frac{1}{8}(k+2)((k+2)^2-m^2)}\Big]
    \ .\nn
\ee
As we did above for the fermionic part of the vector multiplet,
the fermionic term can be diagonalized using the spinor basis $\Psi$.
Then the quadratic operator $\mathcal{O}_{H,f}$ acting on the fermionic field
becomes
\be
	&&\left\{\begin{array}{l}\mathcal{O}_{H,f}\Psi_1 = \big(\!-\!m-1-\Delta+i\alpha(\phi_0)\big)\Psi_1+i\Psi_2 \\
	\mathcal{O}_{H,f}\Psi_2 =-i(k-m)(k+m+4)\Psi_1+\big(m+1+\Delta+i\alpha(\phi_0)\big)\Psi_2
	\end{array}\right. \nn \\
	&&\left\{\begin{array}{l}\mathcal{O}_{H,f}\Psi_3 =\big(\!-\!m+2-\Delta+i\alpha(\phi_0)\big)\Psi_3+i\Psi_4\\
	\mathcal{O}_{H,f}\Psi_4 =-i(k+m)(k-m+4)\Psi_3+\big(m-2+\Delta+i\alpha(\phi_0)\big)\Psi_4\ .
	\end{array}\right.
\ee
Taking into account the degeneracies of $\Psi$, we obtain the fermionic one-loop determinant
\be
	\hspace{-1.2cm}&&{\rm det}_{H,f} \nn \\
	\hspace{-1.2cm}&&=\prod_{\alpha\in root}\left[\prod_{k=0}^\infty\left(k+1+\Delta+i\alpha(\phi_0)
\right)^{\frac{(k+1)(k+2)}{2}}\!\!\prod_{m=-k}^{k-1}\!\!
\left(k(k+4)+2m(\Delta\!-\!1)+(\Delta\!+\!1)^2+\alpha(\phi_0)^2
\right)^{\frac{1}{8}(k+2)((k+2)^2-m^2)}\right.
      \nn \\
   \hspace{-1.2cm} &&\hspace{.3cm} \left.\times\prod_{k=1}^\infty\left(k+2-\Delta-i\alpha(\phi_0)
   \right)^{\frac{(k+1)(k+2)}{2}}\!\!\prod_{m=-k+1}^{k}\!\!\left(k(k+4)+2m\Delta +(\Delta\!-\!2)^2+\alpha(\phi_0)^2\right)^{\frac{1}{8}(k+2)((k+2)^2-m^2)}\right].\quad
\ee
Combining the bosonic and the fermionic determinant, the final one-loop
determinant for the matter hypermultiplet is given by
\be
    \frac{{\rm det}_{H,f}}{{\rm det}_{H,b}} &=&\prod_{\alpha\in root} \prod_{k=0}^\infty\big(k+1+\Delta+i\alpha(\phi_0)\big)^{-\frac{1}{2}(k+1)(k+2)}
    \big(k+2-\Delta+i\alpha(\phi_0)\big)^{-\frac{1}{2}(k+1)(k+2)} \nn \\
    &=& \prod_{\alpha\in root}\prod_{k=1}^\infty
    \big(k-1+\Delta+i\alpha(\phi_0)\big)^{-\frac{k^2-k}{2}}
    \big(k+1-\Delta+\alpha(\phi_0)\big)^{-\frac{k^2+k}{2}}\ .
\ee

\section{Indices and Casimir energies in various dimensions}

In this appendix, we explain how the index captures a quantity similar to the
vacuum Casimir energy of SCFTs on $S^{D-1}\times\mathbb{R}$. It is also interesting
to compare them with the Casimir energy of the dual $AdS_{D+1}$ background.
The Casimir energy is zero in all even dimensional AdS spacetimes \cite{Awad:2000aj},
but is nonzero and proportional to the number of degrees of freedom
of the dual CFT in odd dimensional AdS.

We first study 4d $\mathcal{N}=1$ SCFT. The unrefined superconformal index
contains one fugacity $x$ conjugate to $\epsilon+j$, where $\epsilon$
is the energy and $j$ is the $SU(2)_L\subset SO(4)$ Cartan which rotates $S^3$. Suppose
that the theory contains $n_v$ vector multiplets and chiral multiplets labeled by
$i$ with R-charge $r_i$ for the complex scalar, in $\mathcal{N}=1$ language.
Normally, the index is calculated in a combinatoric way by going to a free theory
limit, or more delicately by going to a UV theory via continuous deformations (such as
RG flows). The elementary quantity is what is called the letter index.
In general, the letter indices for a chiral multiplet with R-charge $r$ and
a vector multiplet are given by \cite{Romelsberger:2005eg,Gadde:2010en}
\begin{equation}
  f_{\rm ch}(x)=\frac{x^{3r/2}-x^{3(2-r)/2}}{(1-x^{3/2})^2}\ ,\ \
  f_{\rm vec}(x)=\frac{2x^3-2x^{3/2}}{(1-x^{3/2})^2}\ ,
\end{equation}
in which the letters are weighted as $x^{\epsilon+j}$.
The full index is given by multiplying to each letter index the character
of the field under the gauge group, then taking the Plethystic exponential.
The final index is obtained by projecting to a gauge singlet.

Although the above combinatoric method captures the information on the
spectrum of BPS states, one would obtain extra multiplicative factor if one
evaluates the index by a path integral. This formally takes the form of
the zero point energy of the vacuum:
\begin{equation}\label{casimir}
  x^{\epsilon_0}\ {\rm with}\ \ \epsilon_0\equiv\frac{1}{2}{\rm tr}\left[(-1)^F
  (\epsilon+j)\right]\ .
\end{equation}
Whenever a free theory description is available (which is the case for many 4d indices
that we can compute), the trace is taken over all the modes of fields. This summation
should be regulated. Since we are considering a supersymmetric path integral which
preserves the SUSY commuting with $\epsilon+j$, a natural (but not compulsory, as
the Casimir energy-like quantity appears to depend on regulator/renormalization, which
is not unique in general without symmetry) regularization is to weight the states with
their $\epsilon+j$ charge. So inserting a factor $x^{\epsilon+j}$ with $x<1$ as
a regulator inside the trace of (\ref{casimir}), one finds that \cite{Kim:2009wb}
\begin{equation}\label{index-casimir}
  \epsilon_0=\frac{1}{2}\lim_{x\rightarrow 1^-}x\frac{d}{dx}f(x)\ ,
\end{equation}
where $f(x)$ is the summation of letter indices of all fields in the theory.

The above regularization is not the usual one which is used to calculate the
vacuum Casimir energy, which is either energy regulator (not $\epsilon+j$) or
the zeta function regularization. For instance, had one been using the energy
regulator, the trace over $j$ would have been zero from rotation symmetry and $\epsilon_0$
would have been the Casimir energy. However, this Casimir energy is not the same as
(\ref{index-casimir}). One can check this for a simple model admitting a free theory
limit. For instance, in the case of free 4d $\mathcal{N}=4$ SYM with $U(N)$ gauge
group, one finds
\begin{equation}\label{N=4-casimir}
  (\epsilon_0)_{\rm true}=\frac{1}{2}\lim_{x\rightarrow 1^-}{\rm tr}
  \left[(-1)^F\epsilon\ x^{\epsilon}\right]=\frac{3N^2}{16}\ ,\ \
  \frac{1}{2}\lim_{x\rightarrow 1^-}{\rm tr}\left[(-1)^F j\ x^{\epsilon}\right]=0
\end{equation}
but
\begin{equation}\label{N=4-index-casimir}
  (\epsilon_0)_{\rm index}=\lim_{x\rightarrow 1^-}{\rm tr}\left[(-1)^F(\epsilon+j)\
  x^{\epsilon+j}\right]=\frac{2N^2}{9}\ .
\end{equation}
In all calculations, we have set the radius of $S^3$ to $1$. In the index,
the former regulator is forbidden by SUSY.
In the latter regularization, $j$ also acquires nonzero value.
If nonzero, all index version of `vacuum charges'
are naturally expected scale like $N^2$.

%

The `index Casimir energy' defined by (\ref{index-casimir}) can be calculated
in general as follows. Defining $x=e^{-\beta}$ and expanding the expression in
(\ref{index-casimir}) for small $\beta$, one obtains
\begin{equation}\label{casimir-expand}
  \frac{x}{2}\frac{df(x)}{dx}=-\frac{2}{3\beta^2}\left(n_v+\sum_i(r_i-1)\right)
  +\frac{1}{8}\left(n_v+\sum_{i}(2(r_i-1)^3-(r_i-1))\right)+\cdots\ .
\end{equation}
Renormalizing away the first divergent term to zero, if the coefficient is nonzero,
the second term would be $\epsilon_0$. It is interesting to compare this with the
$a$ and $c$ central charges of the SCFT, given by \cite{Anselmi:1997am,Intriligator:2003jj}
\begin{eqnarray}
  a=\frac{3}{32}(3{\rm tr}R^3-{\rm tr}R)&=&\frac{3}{32}\left[
  2n_v+3\sum_i(r_i\!-\!1)^3-\sum_i(r_i\!-\!1)\right]\nonumber\\
  c=\frac{1}{32}(9{\rm tr}R^3-5{\rm tr}R)&=&\frac{1}{32}\left[
  4n_v+9\sum_i(r_i\!-\!1)^3-5\sum_i(r_i\!-\!1)\right]\ .
\end{eqnarray}
From these, one finds that $\epsilon_0$ is related to $a$ and $c$ by
\begin{equation}
  \epsilon_0=\frac{2}{9}a+\frac{2}{3}c\ .
\end{equation}
Thus, one finds that $\epsilon_0$ calculated from the index
is always a universal combination of the two central charges.
We also mention in passing that one finds
\begin{equation}
  a-c=\frac{1}{16}\left(n_v+\sum_i(r_i-1)\right)\ ,
\end{equation}
so that the UV divergence given by the first term of (\ref{casimir-expand}) is
zero when $a=c$. The last property holds for CFT models with large $N$ gravity duals
on $AdS_5$ times a smooth 5-manifold.

It is also interesting to compare our index Casimir energy with the proper Casimir
energy. In a QFT having a deformation to a free limit, one can simply calculate it as we
did it above
for the $\mathcal{N}=4$ theory. One can also calculate it from the $AdS_5$ gravity dual
if it exists. We have studied many 4d SCFT with gravity duals, in which case we can study
the true Casimir energy from the gravity dual. We find that the Casimir energy and the index
version of it satisfies a relation
\begin{equation}
  \frac{(\epsilon_0)_{\rm true}}{(\epsilon_0)_{\rm index}}=\frac{27}{32}\ ,
\end{equation}
which is obvious for the $\mathcal{N}=4$ SYM from (\ref{N=4-casimir}), (\ref{N=4-index-casimir}).
This ratio should be universal, as $a=c$ is proportional to $\frac{\ell^3}{G}$ of $AdS_5$
($\ell$: radius, $G$: Newton constant), which in turn is proportional to the Casimir energy.

For 3 dimensional field theories, some of them with $AdS_4$ gravity duals,
the letter indices always contain a factor $\frac{x^{1/2}}{1+x}$ when the scale dimension of
matter fields is canonical. $x$ is again conjugate to $\epsilon+j$, where $j$ is the angular
momentum on $S^2$. From this, one obtains $\lim_{x\rightarrow 1}\frac{df}{dx}=0$.
Therefore, the Casimir energy calculated from the field theory is zero, just like
the true Casimir energy. For non-canonical R-charges with 3d $\mathcal{N}=2$ SUSY,
one finds a factor \cite{Imamura:2011su}
\begin{equation}
  f(x)=\frac{x^{h}-x^{2-h}}{1-x^2}
\end{equation}
for each chiral multiplet, from which one obtains the vanishing Casimir energy as well.

Finally, we study the index Casimir energies of 6d SCFT. We discuss it from the
gravity dual of the large $N$ $(2,0)$ theory. The gravity
Casimir energy on an $AdS_7$ with radius $\ell$ and Newton constant $G$ is given by
\cite{Awad:2000aj}\footnote{\cite{Awad:2000aj} actually calculates the Casimir energy
from the Kerr-$AdS_7$ black hole with single rotation ($j_1\neq 0$, $j_2=j_3=0$ in our notation) by taking out the black hole mass from the stress energy tensor. It still depends
on the rotation parameter, which we turned off to obtain (\ref{AdS7-casimir}).}
\begin{equation}\label{AdS7-casimir}
  -\frac{5\pi^2\ell^4}{128G}\ .
\end{equation}
Using the relation $N^3=\frac{3\pi^2\ell^5}{16G}$, one obtains
\begin{equation}\label{(2,0)-casimir}
 (\epsilon_0)_{\rm gravity}=-\frac{5N^3}{24\ell}\ .
\end{equation}
Like our index Casimir energy, (\ref{(2,0)-casimir}) is negative and scales like $N^3$.


%

\end{document}